\documentclass[11pt]{article}
\usepackage{amsmath, amssymb, amsthm, epsfig}
\newtheorem{theorem}{Theorem}[section]
\newtheorem{Claim}{Claim}[theorem] 
\newtheorem{proposition}[theorem]{Proposition}
\newtheorem{remark}[theorem]{Remark}
\newtheorem{lemma}[theorem]{Lemma}
\newtheorem{corollary}[theorem]{Corollary}

\newtheorem{definition}[theorem]{Definition}

\newcommand{\ggq}{\!\geq\!}

\title{Global spectrum fluctuations for the $\beta$-Hermite and $\beta$-Laguerre ensembles via matrix models}  
\author{Ioana Dumitriu and Alan Edelman}

\begin{document}

\maketitle

\begin{abstract} We study the global spectrum fluctuations for $\beta$-Hermite and $\beta$-Laguerre ensembles via the tridiagonal matrix models introduced in \cite{dumitriu02}, and prove that the fluctuations describe a Gaussian process on monomials. We extend our results to slightly larger classes of random matrices. \end{abstract}

\section{Introduction}

\subsection{The Semicircle Law, deviations and fluctuations, numerically} \label{expe}

\begin{figure}[ht]
\begin{center}\epsfig{figure = 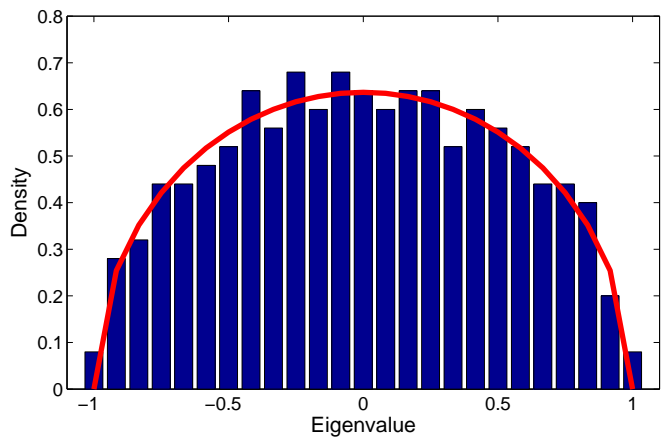, height = 2.3in} \end{center}
\caption{$25$ bin-histogram of the eigenvalues of a $300 \times 300$ symmetric matrix from the real Gaussian distribution, versus the semicircle.} \label{figure1}
\end{figure}

    The most celebrated theorem of random matrix theory, the Wigner semicircle
law \cite{wigner55a, wigner58a}, may be illustrated as in Figure \ref{figure1} by histogramming the eigenvalues of a single random symmetric matrix using the simple MATLAB code (normalization omitted)
\begin{eqnarray*}
& & \tt{  A=randn(n); ~S=(A+A')/sqrt(8* n); ~a = hist(eig(S), [-1:delta:1]);}\\
& & \tt{ bar([-1:delta:1], a/(n *delta))}\\
\end{eqnarray*} 

\vspace{-.3cm}

Every time we run this experiment, we obtain deviations
from the semicircle. The difference from theory is readily explained computationally: 
\begin{itemize} \item  we use a finite matrix size $n$, while the semicircle law is a theorem about the $n=\infty$ limit;
\item histograms bin eigenvalues into boxes of finite width, while the semicircle density is a  continuous function.
\end{itemize}

There can also be numerical error in the experiments due to finite precision
computations and truncation error, but in practice this does not appear to be 
significant.

It is worth noting that the algorithm above is inefficient in two ways: first, it uses the full matrix $A$, rather than the equivalent tridiagonal matrix $H_{\beta,n}$ of Table \ref{tabelu2} (with $\beta = 1$), and it calculates the eigenvalues to obtain the histogram. For more on how to obtain the histogram plot efficiently and without calculating the eigenvalues, see Section \ref{histo}.

To study the next order behavior in the law, for large $n$, we can subtract away the semicircle and multiply by $n$.
%The results of this  numerical experiment appears in Figure \ref{figure2}, 
%where we illustrate the deviation for ??? trials of matrices of size ?????.
%(IF we are lucky...)
%The figure shows that in each interior bin there is a mean that appears to
%be given by a continuous curve that takes on negative values, while on the
%boundaries there is a large positive spike.
 The  next order \emph{average} behavior is what we call
the deviation and it was first computed by Johansson \cite{johansson_clt_herm} to be
\begin{eqnarray} \label{devi}
   DEVIATION  = \frac{1}{4}\delta_{-1}(x) + \frac{1}{4}\delta_1(x) - \frac{1}{2\pi}\sqrt{1-x^2}~.
\end{eqnarray}

This expression for the deviation is the $\beta = 1$ (corresponding to real matrices) instance of the more general $\beta>0$ case (which was also computed by Johansson, and will be explained in the next sections). The $\beta>0$ deviation contains a $\left(\frac{2}{\beta}-1\right)$ multiplicative factor in front of the expression on the right of \eqref{devi}, which disappears for $\beta=1$.

   One can see this deviation result as stating that as $n \rightarrow \infty$ the
eigenvalues are decremented in the interior at a rate that is fastest at the center, pulling the eigenvalues toward the endpoints.

  Upon further examination of the next-order term, one can observe a phenomenon not appearing in the leading order term; there are fluctuations around the mean.  
%These are seen by taking
%multiple trials as we have illustrated in Figure \ref{figure2}.

  Each time we run a trial we can compute this fluctuation.
For example, if we have $25$ bins, the random fluctuation vector $v=(v_1 ,
\ldots, v_{25})^T$ is the difference  between the count in each bin and the
number of eigenvalues predicted by the semicircle plus the deviation.
The entries of $v$ can vary quite wildly, but inner products with
discretized smooth functions result in normal distributions in the 
continuous limit. Specifically, if $f_k$ is the vector of  size $25$ 
consisting of the evaluations of $f(x)=x^k$ on the centers of the bins, 
then the dot products $f_k^Tv$ are heading towards Gaussians with 
covariances $E[ (f_{k_1}^Tv)(f_{k_2}^Tv)]$. 
 
 Precise statements require the size of the matrix 
to go to  infinity and the histogram to melt into a smooth density
function so that the $v$ vector  becomes a Gaussian process.

Denote by $FLUCTUATION( f(x) )$ the random quantity representing the  
limit obtained of $f^Tv$ where, as above, $f$ is the vector of function values
and  $v$ is the vector of histogram differences. For smooth functions, $FLUCTUATION( f(x) )$ converges to a normal distribution; for example, in the limit as $n \rightarrow \infty$, 
\begin{eqnarray} \label{fluctu}
FLUCTUATION(x^k) \sim 2 N(0, \sigma_k^2)~,
\end{eqnarray}
where 
\begin{eqnarray*} \sigma_k^2 = 
\left \{\begin{array}{cl}\frac{k}{2^{2k}} {k -1 \choose \frac{k-1}{2}}^2~,& \mbox{if}~k=1 \mod 2~,\\
 \frac{k}{2^{2k+2}} {k \choose \frac{k}{2}}^2~,& \mbox{if}~k=0 \mod 2~. \end{array} \right .
\end{eqnarray*}

For general $\beta$, the right side of \eqref{fluctu} gains a multiplicative factor of $\frac{1}{\beta}$.

The limit of the entry $(k_1, k_2)$ of the covariance matrix becomes the covariance between $FLUCTUATION(x^{k_1})$ and $FLUCTUATION(x^{k_2})$ (see Theorem \ref{main_herm} with $\beta = 1$).

\subsection{$\beta$-Hermite and $\beta$-Laguerre ensembles} \label{intro}

%Random matrix theory is a field ever growing in importance because of three reasons: the wide aplicability of the formulas and methods, the ability to provide software tools, and the fundamental ability to provide exact closed form or asymptotic expressions. The last reason should not be understated, the underlying apparatus has the benefit of being so highly connected to so many areas of mathematics that no one field can ``own'' random matrix theory, though many can contribute. The mathematics falls in that very nice place such that exact analytical or asymptotic expressions are derivable, but not trivially.  Thus current research continues to provide new exact formulas at a rapid pace.
This paper studies  deviations and fluctuations in a wider
context than real symmetric  matrices with semicircular asymptotic density:
We consider Hermite and Laguerre matrices with general parameter $\beta>0$. For a great reference for these ensembles, see Forrester's upcoming book \cite{Forrester_book}.

Classical finite random matrix theory considers the study of eigenvalue ensembles with joint density 
\[
f(x_1, \ldots, x_n) = c_{w, n}~ \prod_{i<j} |x_i - x_j|^{\beta} \prod_{i=1}^n w(x_i)~,
\]
with $w$ a scalar weight function on an interval $I$. This interval may be a subinterval of the real line, or the unit circle in the complex plane; other possibilities have been considered, too, and generalizations are easily conceived. A good reference for these formulae can be found in Mehta's book \cite{mehta_book}. 

Some of the most studied eigenvalue ensembles have Hermite, Laguerre, and Jacobi weight functions on the real line, or uniform weight on the unit circle. In this paper we will be examining the ensembles with Hermite and Laguerre weights on the real line (respectively, half-line); see Table \ref{tabelu1}. 

\begin{table}[ht] 
\caption{Random matrix ensembles with eigenvalue distribution proportional to $f(x_1, \ldots, x_n) ~\varpropto~ \prod_{i \neq j} |x_i - x_j|^{\beta} ~\prod w(x_i)$ with $w$ defined on an interval $I \subseteq \mathbb{R}$. MANOVA stands for Multivariate ANalisys Of VAriance.} \label{tabelu1}

\vspace{.5cm}

\begin{tabular}{l|c|c|c|c}
\textit{Name} \!\!& \textit{parameters} & $I$ & $w$  & \textit{Historical} \\
& & & & \textit{name/constraints} \\
\hline 
\hline
\!\!\textbf{Hermite} \!\!& {\mathversion{bold} $\beta>0$} &$\mathbb{R}$ & $w(x) = e^{-x^2/2}$ & Gaussian \\
& & & & $\beta = 1,2,4$ \\
& & & & \\
\hline 
\hline
\!\!\textbf{Laguerre}\!\!& {\mathversion{bold}$\beta>0$} & $[0, \infty)$ &$w(x) = x^{p} e^{-x/2}$ &  Wishart \\
& $~a>(n-1) \frac{\beta}{2}$ & &$p=a- (n-1)\frac{\beta}{2}-1$ & $\beta=1,2$ \\
&& &&$a = m \frac{\beta}{2}$, $m \in \mathbb{N}$  \\
& & & & \\
\hline 
\hline
\!\!\textbf{Jacobi} \!\!&  {\mathversion{bold}$\beta>0$} & $[0, 1]$ & $w(x) = x^{p_1} (1-x)^{p_2}$ & MANOVA \\
& $a_1, a_2>(n-1) \frac{\beta}{2}$ & &$p_1 = a_1- (n-1)\frac{\beta}{2}-1$ & $\beta = 1,2$ \\
& & & $p_2 =a_2- (n-1)\frac{\beta}{2}-1$&   $a_1 = m_1 \frac{\beta}{2}$, $m_1 \in \mathbb{N}$  \\
& & & & $a_2 = m_2 \frac{\beta}{2}$, $m_2 \in \mathbb{N}$\\
\end{tabular}
\end{table}

For more references on Gaussian ensembles, see \cite{mehta_book}; for Wishart and MANOVA ensembles, see \cite{muirhead82a}; for Hermite, Laguerre, and Jacobi ensembles, see \cite{Forrester_book}.

For three particular values of $\beta$, namely $1,2,$ and $4$, these ensembles have been studied since the birth of the field, as the Gaussian real, complex, and quaternion ensembles (Hermite with $\beta=1,2,4$) of nuclear physics (\cite{wigner55a}, \cite{wigner58a}, \cite{dyson_3fold}, \cite{arnold71a}). Similarly, the Wishart real and complex (Laguerre with $\beta = 1,2$ and some restrictions on the Laguerre parameter) matrices emerged from the world of statistical multivariate analysis (\cite{wishart28}, \cite{constantine_noncentral}, \cite{james64a}, \cite{jonsson82a}). 

The parameter $\beta$ (making the connection to the Boltzmann factor of statistical physics) is seen by some communities (e.g. statistical mechanics) as an inverse temperature, or repulsion strength, of the ensemble of eigenvalues (the higher the $\beta$, the more separated the eigenvalues). It also has the advantage of the easy mnemonic of $1, 2$ and $4$ corresponding to real, complex, and quaternion entries in the matrix models. However, some communities (like algebraic combinatorics) consider a different parameter, $\alpha = 2/\beta$, which tends to simplify certain formulas (\cite{MacDonald_book}, \cite{Stanley_jacks}). In this paper we will use both notations, for convenience, and make sure that the reader is informed when changes take place.

The reason for the attractivity and success that the study of Gaussian 
orthogonal, unitary, and symplectic (Hermite with $\beta = 1,2,4$), and the Wishart real and complex (Laguerre with $\beta=1,2$), ensembles have enjoyed lies in the existence of matrix models with real, complex, and quaternion entries (see \cite{mehta_book}, \cite{muirhead82a}). These models have not only originated the study, but have also allowed for a relatively thorough analysis of the finite and asymptotical eigenstatistics (statistical properties of the eigenvalues). 

One of the developments in the study of arbitrary $\beta$-Hermite, -Laguerre, and -Jacobi ensembles is the introduction of general \emph{real} matrix models (see \cite{dumitriu02}, \cite{killip-nenciu}, \cite{sutton05}). For every $\beta$, there are simple \emph{real tridiagonal matrices} which model the corresponding eigenvalue distributions given by Table \ref{tabelu1}. For the $\beta$-Hermite and -Laguerre ensembles, we present these forms in Table \ref{tabelu2}. These matrix forms allow for efficient Monte Carlo experiments and an alternative representation for the study of eigenstatistics in the general $\beta$ case. 

This paper contains one example of such analysis. With the help these matrix models, we compute asymptotical global spectrum fluctuations for the $\beta$-Hermite and -Laguerre ensembles. For the latter, these results are new in the general $\beta$ context; the global spectrum fluctuations of complex Wishart matrices ($\beta$-Laguerre with $\beta =2$) were studied by Speicher et al. \cite{speicher_new}, and before by Cabanal-Duvillard \cite{cabanal_duvillard}. Johansson's more extensive study \cite{johansson_clt_herm} covers our results for the $\beta$-Hermite ensembles. 
 
 To prove our theorems, we use a very diverse set of methods and techniques from Jack Polynomial theory, special functions, perturbation theory, and combinatorial path-counting.
Some of the techniques we used in studying the traces of powers of random matrices have been inspired by the work of Soshnikov and Sinai \cite{sosh_sinai}, \cite{sosh_sinai2}, and by the study of traces of unitary random matrices by Diaconis and Shahshahani \cite{diaconis93a}.

\begin{table}[ht] 
\caption{\textbf{Unscaled and scaled} tridiagonal matrix models for the $\beta$-Hermite and $\beta$-Laguerre ensembles with any $\beta>0$, $n \in \mathbb{N}$, $a \in \mathbb{R}$, and $a > \frac{\beta}{2} (n-1)$.} \label{tabelu2}

\begin{center} \begin{tabular}{|l||c|} \hline & \\ $\begin{array}{c}
\mbox{\textbf{Hermite} matrix,} \\ \textbf{unscaled} \end{array}$ & {\huge $H_{\beta,n}$} $
\sim \frac{1}{\sqrt{2}} \left( \begin{array}{ccccc} N(0, 2) & \chi_{(n-1) \beta} & & & \\
\chi_{(n-1) \beta} & N(0, 2)  & \chi_{(n-2) \beta} & & \\ & \ddots & \ddots & \ddots & \\
& & \chi_{2\beta} & N(0,2) & \chi_{\beta} \\ & & & \chi_{\beta} & N(0,2) \end{array}
\right)$ \\ & \\ \hline & \\ $\begin{array}{c}\mbox{\textbf{Hermite} matrix,} \\ \textbf{scaled} \end{array}$ & {\huge $\tilde{H}_{\beta,n} ~=~ \frac{1}{\sqrt{2n \beta}} ~H_{\beta,n}$} \\  &  \\
& \\ \hline \hline  & \\ 
$\begin{array}{c} \mbox{\textbf{Laguerre} matrix,} \\ \textbf{unscaled} \end{array}$ & {\huge $L_{\beta,n}^{a}$} $=
B_{\beta,n}^{a} (B_{\beta,n}^{a})^{T}$, where \\  & \hspace{1cm} $B_{\beta,n}^{a} \sim \left(
\begin{array} {cccc} \chi_{2a} & & & \\ \chi_{\beta (n-1)} & \chi_{2a-\beta} & & \\ &
\ddots & \ddots & \\ & & \chi_{\beta} & \chi_{2a-\beta(n-1)} \end{array} \right)$ \\ & \\ \hline & \\ 
$\begin{array}{c} \mbox{\textbf{Laguerre} matrix,} \\ \textbf{scaled} \end{array}$ & {\huge $\tilde{L}_{\beta,n}^{a} ~=~ \frac{\gamma}{n \beta} ~L_{\beta,n}^{a}$} \\ & \\
\hline \end{tabular} \end{center} 

\end{table}

\subsection{Statements of results}

 Among the asymptotical eigenstatistics one may distinguish two classes of properties: local (like the scaled distributions and fluctuations of extremal eigenvalues, or like the level spacing distribution, i.e., the distance between neighboring eigenvalues) and global (like the limiting level density, i.e. the distribution of a random eigenvalue, and fluctuations thereof). The term global tends to refer to a property that involves all or a significant portion of the eigenvalues, while local tends to refer to a property that occurs near an individual or a constant number of eigenvalues. Among the more famous local properties we enumerate the level spacings for the Gaussian ensembles (see \cite{mehta_book}, \cite{tracy_widom_fredholm}) and the extremal (largest, corresponding to ``soft edge'', smallest, to ``hard edge'') eigenvalue asymptotics (see \cite{tracy_widom_largest}, \cite{tracy_widom_1_4}, \cite{johansson_wishart}, \cite{johnstone}). All these results relate to real, complex, or quaternion matrices; in the category of results relating to general $\beta$, we have to mention recent work by Desrosiers and Forrester \cite{Forrester_corrections}, where they analyze the asymptotical corrections to the eigenvalue density, and, for $\beta \in 2 \mathbb{N}$, obtain the expected $O(n^{2/3})$ order in the fluctuation of the largest eigenvalue in the case of both $\beta$-Hermite and -Laguerre ensembles.

For $n$ (the size of the ensembles) finite, the $n$ eigenvalues may be considered as $n$ fluctuating particles. Roughly speaking, the larger the $\beta$, the less fluctuation there is in the particles (hence $\beta$ is seen as an inverse temperature).  As $\beta \rightarrow \infty$ the particle positions  behave
like multivariate normals with variance $O(1/\beta)$ and means located at the roots of a Hermite (respectively Laguerre) polynomial (see \cite{dumitriu04a}).  

For a fixed $\beta$, as $n \rightarrow \infty$ and, in the case of Laguerre ensembles, $2a/(n \beta) \rightarrow \gamma$, the particles have an
emerging (global) level density which is obeys a simple law (Wigner's semicircle law \cite{wigner58a} for the Hermite ensembles, Mar\v{c}enko-Pastur laws \cite{marcenko67a} for the Laguerre ensembles).
 The roots of the Hermite, respectively Laguerre, polynomial have this same asymptotical density -- the fluctuations do not change the asymptotics, as they are on a smaller scale.

In \cite{dumitriu03th}, we have proved convergence almost surely (as $n \rightarrow \infty$) of the asymptotical eigenvalue distribution of the $\beta$-Hermite ensemble to the semicircle distribution $S$ with density $\frac{2}{\pi} \sqrt{1 - x^2}$, and of the asymptotical eigenvalue distribution of the $\beta, a$-Laguerre ensemble (with $n \beta/(2a) \rightarrow \gamma \leq 1$)  to the Mar\v{c}enko-Pastur $E_{\gamma}$ distribution with density $ \frac{1}{2 \pi \gamma} \frac{\sqrt{(x-(\sqrt{\gamma}-1)^2)((\sqrt{\gamma}+1)^2-x)}}{x}$. We recall that convergence almost surely is stronger (and implies) convergence in distribution, a.k.a. convergence of moments.

We examine here the distribution of the statistic $\sum\limits_{i=1}^n f(\lambda_i)$, where $f$ is a function of the \emph{scaled} eigenvalues $\lambda_i$. The scaling is $\lambda \rightarrow \sqrt{2n \beta} \lambda$ for the $\beta$-Hermite ensembles and $\lambda \rightarrow n \beta/\gamma \lambda$ for the $\beta$-Laguerre ensembles (see Table \ref{tabelu2}).

General $\beta$ results for this kind of statistic can be found in \cite{BPS}, \cite{johansson_clt_herm}; this or similar linear statistics have been considered also in \cite{forrester_global} (for unitary matrices), \cite{silverstein03} (for Wishart matrices), and, heuristically, in \cite{politzer}. Another path of interest is represented by asymptotical large deviations from the density (spectral measure); we mention the results of \cite{cabanal_duvillard}, \cite{benarous_ld}, \cite{guionnet02a} (which also covers global fluctuations for Wishart matrices) , \cite{guionnet02b}, \cite{guionnet03a} (which covers moderate deviations).

For the linear statistic $\sum_{i=1}^n f(\lambda_i)$, the Wigner and Mar\v{c}enko-Pastur laws for the $\beta$-Hermite and $\beta$-Laguerre ensembles state that for any ``well-behaved'' function $f$,
\begin{eqnarray} \label{wig}
\frac{1}{n} ~\sum_{i=1}^n f(\lambda_i) & \rightarrow & \frac{2}{\pi}\int_{-1}^{1} f(t) \sqrt{1-t^2}~dt~, ~\mbox{respectively,} \\
\label{m-p} \frac{1}{n} ~\sum_{i=1}^n f(\lambda_i) & \rightarrow & \frac{1}{2 \pi \gamma} \int_{(\sqrt{\gamma}-1)^2}^{(\sqrt{\gamma}+1)^2} f(t) ~\frac{\sqrt{(t-(\sqrt{\gamma}-1)^2)((\sqrt{\gamma}+1)^2-t)}}{t}~dt~,
\end{eqnarray}
where the convergence in the above is almost surely\footnote{Such a result is sometimes given the name of ``Strong Law of Large Numbers''; if convergence is only in distribution, the name becomes ``Law of Large Numbers''.}. 

Examining the fluctuations from these laws takes us one step further. For $f$ a polynomial, we prove that once we subtract the expected average over the limiting level density (i.e. the right hand sides of \eqref{wig} and \eqref{m-p}), the rescaled resulting quantity tends asymptotically to a normal distribution with mean and variance depending on $f$. In other words, once the semicircle or Mar\v{c}enko-Pastur distributions are subtracted, the fluctuations in the statistic $\sum\limits_{i=1}^n f (\lambda_i)$ tend asymptotically to a Gaussian process on polynomials $f$.

%By analogy we consider Gaussian processes $W(x)$.  These have the property \marginpar{uh-huh!}\footnote{This is mathematically iffy, gotta reformulate.} that $\int h(x)W(x)dx$ are univariate Gaussians for ``reasonable'' functions $h$. Hhow ``reasonable'' $h$ has to be depends on the covariance matrix of this Gaussian process (with respect to some basis). For example, consider the space of polynomials $\mathbb{R}[x]$. If the variables $t_i = \int x^i W(x) dx$, for all $i \in \mathbb{N}$, have the property that $C_{ij} = Cov(t_i, t_j) < \infty$, for all $i, j \in \mathbb{N}$, it follows that the Gaussian process $W$ can be defined on the class of polynomials $\mathbb{R}[x]$. 

How much more we can extend the class of functions that this process is well-defined on depends on the entries of the covariance matrix $C = \{C_{ij}\}$ (which can be expressed in any polynomial basis). One would be tempted to believe that a Gaussian process $W$ defined on polynomials could,  in principle, be extended to a class of continuous functions $h(x)$ with the property that, given a sequence of polynomials $p_n(x) \rightarrow h(x)$ in some norm, given $v_n = Var_{W}(p_n)$, $v_n \rightarrow \tilde{v} < \infty$, such that $\tilde{v} = Var_{W}(h)$.

%We can define an inner product of functions that carry all the covariance
%information:
%$$<h,h> \equiv $E(\int h(x)w(x))^2.$

\begin{definition}
Let $\gamma \in [0,1]$ be a real parameter, and let $a = (\sqrt{\gamma}-1)^2$, $b = (\sqrt{\gamma}+1)^2$. We define the following two measures:
\begin{eqnarray*}
\mu_{H}(x) & := & \left \{ \begin{array}{cl} \frac{1}{4} \delta_1(x) + \frac{1}{4}\delta_{-1}(x) - \frac{1}{2\pi} \frac{1}{\sqrt{1 - x^2}}, & \mbox{if $x \in [-1, 1]$}~, \\ 0~, & \mbox{otherwise.} \end{array} 
\right . ~, \\
\mu_{L}^{\gamma}(x) & := &  \left \{ \begin{array}{cl} \frac{1}{4} \delta_{b}(x) + \frac{1}{4} \delta_a(x) - \frac{1}{2 \pi} \frac{1}{\sqrt{(x-a)(b-x)}}~, & \mbox{if $x \in [a, b]$}~, \\ 0~, & \mbox{otherwise~.} \end{array} 
\right .
\end{eqnarray*}
\end{definition}

\begin{theorem} \label{main_herm}
Let $\tilde{H}_{\beta, n}$ be a scaled matrix from the $\beta$-Hermite ensemble of size $n$, with (scaled) eigenvalues $(\lambda_1, \ldots, \lambda_n)$, and let $k \geq 1$ be a positive integer. For all $i =1, \ldots, k$, let
\begin{eqnarray*}
X_i & = & tr((\tilde{H}_{\beta,n})^i) - n ~\frac{1}{4^{i/2} \left( \frac{i}{2}+1 \right) } {i \choose \frac{i}{2}} \delta_{(i \!\!\!\!\mod 2), 0} - \left(\frac{2}{\beta} -1 \right) \int_{-1}^{1} t^i \mu_H(t)~dt,\\
& \equiv & \sum_{j=1}^n \lambda_j^i ~-~ n ~\frac{2}{\pi} \int_{-1}^{1} t^i \sqrt{1-t^2} ~dt ~-~ \left(\frac{2}{\beta} -1 \right) \int_{-1}^{1} t^i \mu_H(t)~dt~~.
\end{eqnarray*}
Let $(Y_1, Y_2, \ldots, Y_k)$ be a centered multivariate Gaussian with covariance matrix 
\begin{eqnarray} \label{cov_mat}
\mbox{Cov}(Y_i, Y_j) = \left \{ \begin{array}{ll} \frac{1}{2^{i+j}} \frac{2ij}{i+j} {i-1 \choose \frac{i-1}{2}} {j-1 \choose \frac{j-1}{2}}~,& \mbox{if}~~i=j=1 \mod 2~; \\
 \frac{1}{2^{i+j+2}} \frac{2ij}{i+j} {i \choose \frac{i}{2}} {j \choose \frac{j}{2}}~,&\mbox{if}~~i=j=0 \mod 2~; \\
0~, & \mbox{otherwise.} \end{array} \right .\end{eqnarray}

Then, as $n \rightarrow \infty$,
\[
(X_1, X_2, \ldots, X_k) \Rightarrow \sqrt{\frac{2}{\beta}} ~(Y_1, Y_2, \ldots, Y_k)~.
\]
\end{theorem}

\begin{remark} 
For any size $k$, if we add a first row and a first column of zeros to the covariance matrix \eqref{cov_mat}, the resulting $(k+1) \times (k+1)$ matrix $C_k$ has a scaled Cholesky decomposition as $C_k = T_k D_k T_k^{T}$, where $D_k$ is the diagonal matrix having on the diagonal the vector $(0, 1, 2, 3, \ldots, k)$, and $T_k$ has an interpretation as the change-of-base matrix in the space of univariate polynomials from monomials basis to the Chebyshev polynomials basis. One can also look at the infinite version $C_{\infty} = T_{\infty} D_{\infty} T_{\infty}^T$.   
\end{remark}

\begin{remark}
This Gaussian process is extended to a larger class of continuous functions in \cite{johansson_clt_herm}. Johansson conjectured that the regularity conditions imposed were purely technical, and that in fact the correct condition should be that the function $h$ admits a Fourier-like expansion in the Chebyshev basis, which we write as an (infinite) vector $\vec{h}$, such that $\vec{h}^{T} D_{\infty} \vec{h}$, the variance of $h$ under the process, is finite. This conjecture is equivalent to saying that the process could be extended to any function $f$ such that  $\vec{f}^{T} C_{\infty} \vec{f}$ is finite (since we have chosen the monomial basis as the representation, rather than the Chebyshev polynomial basis).
\end{remark}

\begin{theorem} \label{main_lag}
Let $\tilde{L}_{\beta, n}^{a}$ be a scaled matrix from the $\beta$-Laguerre ensemble of parameter $a$ and size $n$, with (scaled) eigenvalues $(\lambda_1, \ldots, \lambda_n)$, and let $k \geq 1$ be a positive integer. Assume that $n \beta/(2a) \rightarrow \gamma \leq 1$, and let $\gamma_{min} = (\sqrt{\gamma}-1)^2,~\gamma_{max} = (\sqrt{\gamma}+1)^2$. For all $i =1, \ldots, k$, let
\begin{eqnarray*}
X_i & = & tr((\tilde{L}^{a}_{\beta,n})^i) - n ~\sum_{r=0}^{i-1} \frac{1}{r+1} {k \choose r} {k-1 \choose r} \gamma^r - \left(\frac{2}{\beta} -1 \right) \int_{\gamma_{min}}^{\gamma_{max}} \!\!t^i \mu_L^{\gamma}(t)~dt,\\
& \equiv & \sum_{j=1}^n \lambda_j^i ~-~ n ~\frac{1}{2\pi \gamma} \int_{\gamma_{min}}^{\gamma_{max}} \!\!t^i \sqrt{(t-\gamma_{min})(\gamma_{max} -t)} ~dt ~-~ \left(\frac{2}{\beta} -1 \right) \int_{\gamma_{min}}^{\gamma_{max}} \!\!t^i \mu_L^{\gamma}(t)~dt~~.
\end{eqnarray*}
Let $(Y_1, Y_2, \ldots, Y_k)$ be a centered multivariate Gaussian with covariance matrix 
\begin{eqnarray} \label{cov_mat_lag}
\mbox{Cov}(Y_i, Y_j) = T_D(i,j) +T_S(i,j)~,
\end{eqnarray}
where 
\[
T_D(i,j) = \sum_{q=1}^{i+j-1} (-1)^{q+1} \gamma^{i+j-q} \frac{{i+j \choose q}}{i+j} \sum_{l=q+1}^{i+j} \frac{(-1)^{l}}{{i+j -1 \choose l-1}} \sum_{\begin{array}{c} r+s=l \\ 1 \leq r \leq i \\ 1 \leq s \leq j \end{array}} rs {i \choose r}^2 {j \choose s}^2~,
\]
and 
\[
T_S(i,j)= \sum_{q=0}^{i+j-2} (-1)^q \gamma^{i+j-q} \frac{{i+j \choose q}}{i+j} \sum_{l=q}^{i+j-2} \frac{(-1)^{l}}{{i+j -1 \choose l}} \sum_{\begin{array}{c} r+s=l \\ 0 \leq r \leq i-1 \\ 0 \leq s \leq j-1 \end{array}} (i-r)(j-s) {i \choose r}^2 {j \choose s}^2~.
\]

Then, as $n \rightarrow \infty$,
\[
(X_1, X_2, \ldots, X_k) \Rightarrow \sqrt{\frac{2}{\beta}} ~(Y_1, Y_2, \ldots, Y_k)~.
\]
\end{theorem}

\begin{remark} For any size $k$, if we add a first row and a first column of zeros to the covariance matrix \eqref{cov_mat_lag}, the resulting matrix $C$ should admit a scaled Cholesky decomposition of the form $C_k = T_k D_k T_k^{T}$, with $D_k$ being the diagonal matrix with diagonal entries $\{j \gamma^{-j}\}_{0 \leq j \leq k}$, and $T_k$ being the change-of-basis matrix in the space of polynomials from monomial basis to the \emph{shifted Chebyshev} polynomials of the first kind (as defined by Cabanal-Duvillard \cite{cabanal_duvillard} and used in \cite{speicher_new}). Note that the constant $c$ in \cite{speicher_new} is our $1/\gamma$. It may also be useful to look at $C_{\infty} = T_{\infty} D_{\infty}T_{\infty}^{T}$.\end{remark}

%For the Hermite ensembles, Johansson's results [] subsume Theorem \ref{main_herm}. Theorem \ref{main_lag}, in the general $\beta$ form is original. 

\begin{remark}
It is worth noting that results like Theorems \ref{main_herm} and \ref{main_lag}, where one averages over a set of quantities, then subtracts the mean and scales by the variance to obtain a limiting Gaussian, are sometimes called \emph{\textbf{central limit theorems}} (see for example \cite{silverstein03} and \cite{sosh_sinai}).  Free Probability uses this term as well, see for example \cite{speicher05a}, in a different context, namely, to express the fact that averaging over random matrices creates an eigenvalue distribution that approaches the semi-circular law. Both uses of the term ``central limit theorem'' draw different parallels to the classical case. 
\end{remark}

Our approach to proving Theorems \ref{main_herm} and \ref{main_lag} 
consists of computing the first-order deviation from the mean (in Section \ref{mean}), showing that the centered process is Gaussian on monomials (by the method of moments), and computing the covariance matrices (in Section \ref{gauss}). 

Finally, in Section \ref{gen}, we generalize our approach to two different classes of random matrices.

\section{Deviation from the semicircle and Mar\v{c}enko-Pastur laws} \label{mean}

\subsection{Dependence on $\beta$: symmetric functions and the ``palindrome'' effect} \label{palindrome}

As stated in the introduction, we are interested in computing the deviation to the semicircle and Mar\v{c}enko-Pastur laws (denoted below by $LAW(\infty)$, as opposed to the $LAW(n)$, which is the level density for finite $n$). These deviations have the form 
\[
LAW(n) \sim LAW(\infty) + \frac{1}{n} \left ( 1-\frac{2}{\beta}  \right) DEVIATION( \beta = \infty) + o \left ( \frac{1}{n} \right)~,
\]
as $n \rightarrow \infty$.

By integrating the above against $x^k$, we can write this in the moment form
\[
moment_k(n) = moment_k(\infty) + \frac{1}{n} \left ( 1-\frac{2}{\beta} \right) moment_k(DEVIATION(\beta= \infty)) + o \left ( \frac{1}{n} \right)~,
\]
again as $n \rightarrow \infty$.

We mention two interesting points, the first of which we prove in this section: \begin{enumerate} \item The factor $\frac{2}{\beta} - 1$ can be obtained from a symmetry principle alone. It is a direct consequence of Jack Polynomial theory that the coefficient of $1/n^j$ in $moment_k(n)$ is a \emph{palindromic} polynomial (we define ``palindromic'' below) in $-\frac{2}{\beta}$, and from the tridiagonal matrix models it follows that the degree of this polynomial is $j$; thus when $j=1$ this polynomial must be a multiple of $\frac{2}{\beta} - 1$. Mathematically, this is significant because because in order to study the
deviations, it is sufficient to  then study the non-random
case, $\beta=\infty$. In summary, the powerful Jack Polynomial theory
allows us to take a complicated random matrix
problem and  reduce it to an exercise on
the properties of   univariate Hermite and Laguerre polynomials.

\item  With the Maple Library \emph{MOPs} \cite{dumitriu04b}, we can compute
symbolically  the exact values of $moment_k(n)$ for small values of $k$, \emph{ as a function of $n$ and $\beta$}. In other words, while this paper concerns itself with the constant and $O(1/n)$ behavior, it is worth remembering
that higher order terms are in principle available to us.
\end{enumerate}

%To study the dependence on $\beta$ in the mean of the statistic $\sum_{i} f(\lambda_i)$, we will make use of a deep symmetry embedded in the integral of the moments 
%of the trace (i.e. the expectation), which we call the ``palindrome" 
%effect. This symmetry may be first appreciated by considering concrete 
%formulas like the ones we present below. 

To make notation a bit clearer, we have used the $\alpha = 2/\beta$ in the below; we also recall the scaled matrices $\tilde{H}_{\beta, n}$ and $\tilde{L}_{\beta, n}^{a}$, from Section \ref{intro}. The following are the first three non-trivial moments for the traces of the scaled $\beta$-Hermite and $\beta$-Laguerre matrices with $a =n \beta/(2 \gamma) = n/(\alpha \gamma)$. We omit the $n$ and $a$ in the notation for reasons of space.
\begin{eqnarray*} 
\!\!\!\!\frac{1}{n} ~E [ \mbox{tr}(H_{2/\alpha}^2) ] & = & \frac{1}{4} +  \frac{\alpha-1}{4n},~\\ 
\!\!\!\!\frac{1}{n} ~E [ \mbox{tr}(H_{2/\alpha}^4) ] & = & \frac{2}{16} + \frac{5\alpha-5}{16n} + \frac{3 \alpha^2 - 5 \alpha 
+3}{16 n^2},~\\ 
\!\!\!\!\frac{1}{n} ~E [ \mbox{tr}(H_{2/\alpha}^6) ] & = & \frac{5}{64} + 
\frac{11\alpha-11}{32n} + \frac{16 \alpha^2 - 27 \alpha 
+16}{32n^2} + \frac{15\alpha^3 - 32 \alpha^2 +32 \alpha - 
15}{64n^3},~ \\
\!\!\!\!\frac{1}{n} ~E [ \mbox{tr}(L_{2/\alpha}) ]  & = & 1~, \\
\frac{1}{n} ~E [ \mbox{tr}(L_{2/\alpha}^2) ] & = & (1+\gamma) + \frac{\gamma( \alpha - 1)}{n} ~,\\
\frac{1}{n} ~E [ \mbox{tr}(L_{2/\alpha}^3) ]  & = & (1 +3 \gamma + \gamma^2) + \frac{3 \gamma ( \gamma+1) ( \alpha - 1)}{n} + \frac{\gamma^2(2 \alpha^2 - 3 \alpha +2)}{n^2}~.
\end{eqnarray*}

Note that the $O(1)$ terms in the above correspond to the second, fourth, and sixth moments of the semicircle, in the Hermite case, respectively, to the first, second, and third moments of the Mar\v{c}enko-Pastur distributions in the Laguerre case; these are Catalan numbers (scaled down by powers of $4$ because of the semicircle $[-1, 1]$ normalization), respectively 
 Narayana polynomials in $\gamma$. 

The $O(1/n)$ terms, the moments of the deviation, are always  multiplied by $\alpha-1$, while the other coefficients of the negative powers of $n$ in the above are ``palindromic polynomials'' of $(-\alpha)$; we recall the definition below.

\begin{definition} A classical ``palindromic polynomial'' is defined by the fact that its list of coefficients, is the same whether read from beginning to end or from end to beginning.
\end{definition}

\begin{remark} An odd-degree palindromic polynomial in $x$ is a multiple of $(x+1)$. \end{remark}

%The combinatorialist will immediately notice that zero-order coefficient in the list of expectations above is always the Catalan number corresponding either to half the exponent in the Hermite case, or to the exponent itself in the Laguerre one.

%The expectation formulas, which have a clear (albeit slightly complicated to explain) symmetry, are not accidental. They are a consequence of the symmetric properties of a class of well-studied, though not very well-known, class of functions: the Jack polynomials. 

To prove that the dependence of the first-order term in the deviation is indeed a multiple of $\alpha - 1$, i.e., of $\frac{2}{\beta}-1$, we will use elements of Jack polynomial theory, and also a stronger form of a 
duality principle proved in \cite{dumitriu03th}. 

We introduce below two notational conventions to be used throughout the rest of the paper. 

\begin{definition} 
We denote by $E_{\delta}^{H}[P(x_1, \ldots, x_s)]$, respectively $E_{\delta, a}^{L}[P(x_1, \ldots, x_s)]$, the expectations of the polynomial $P$ over the \textbf{scaled} $2/\delta$-Hermite, respectively $2/\delta, a$-Laguerre, ensembles of size $s$. 

We denote by $\mathcal{E}_{\delta}^{H}[P(x_1, \ldots, x_s)]$, respectively $\mathcal{E}_{\delta, a}^{L}[P(x_1, \ldots, x_s)]$, the expectations of the polynomial $P$ over the \textbf{unscaled} $2/\delta$-Hermite, respectively $2/\delta, a$-Laguerre, ensembles of size $s$. 
\end{definition}

Let $\mathbb{R}[x_1, \ldots, x_n]$ be the space of symmetric polynomials in $n$ variables (by symmetric we mean invariant under any permutation of the variables). A homogeneous basis for this vector space is a set of linearly independent, symmetric and homogeneous polynomials which generate $\mathbb{R}[x_1, \ldots, x_n]$. One such basis is given by the power-sum functions, defined multiplicatively below. For reference, see \cite{EC2} and \cite{MacDonald_book}.

\begin{definition}
Let $\lambda \equiv (\lambda_1, \lambda_2, \ldots, \lambda_n)$ denote an ordered partition $(\lambda_1 \ggq \lambda_2 \ggq \ldots \ggq \lambda_n$). We define the power sum functions by
\begin{eqnarray*}
p_{\lambda_i} & = & \sum_{j=1}^n x_j^{\lambda_i}~, ~~~~\mbox{and} \\
p_{\lambda}  & = & p_{\lambda_1} p_{\lambda_2} \ldots p_{\lambda_n}~.
\end{eqnarray*}
\end{definition}

The Jack polynomials $J_{\lambda}^{\alpha}$ constitute a parameter-dependent (the parameter being usually denoted by $\alpha$) class of orthogonal multivariate polynomials; they are indexed by the powers of the highest-order term $\lambda$ (in lexicographical ordering).

Throughout this section, we will think of the parameter $\alpha$ as a sort of inverse to $\beta$ (recall that we have denoted $\alpha = 2/\beta$). 

The Jack polynomials allow for several equivalent definitions (up to certain normalization constraints). We will work here with definition \ref{oo}, which arose in combinatorics. We follow Macdonald's book \cite{MacDonald_book}.

\begin{definition} \label{oo} The Jack polynomials $J_{\lambda}^{\alpha}$ are orthogonal with respect to the inner product defined below on power-sum functions 
\[
\langle p_{\lambda}, p_{\mu} \rangle_{\alpha} = \alpha^{l(\lambda)} z_{\lambda} \delta_{\lambda \mu},
\]
where $z_{\lambda} = \prod\limits_{i=1}^{l(\lambda)} a_i!i^{a_i}$,  $a_i$ being the number of occurrences of $i$ in $\lambda$. In addition, the coefficient of the lowest-order term in $J_{\lambda}^{\alpha}$, which corresponds to the partition $[1^{|\lambda|}] \equiv (1,1,\ldots,1)$ (of length $|\lambda|$), is $|\lambda|!$. 
\end{definition}

From now on, we will use the notations $I_n$ for the vector of $n$ ones and we will refer to the quantity $J_{\kappa}^{\alpha}(x_1, \ldots, x_n)/J_{\kappa}^{\alpha}(I_n)$ as the \emph{normalized} Jack polynomial.

To prove our results, we will need the two lemmas  below, the first of which is a stronger variant of the duality principle proved in \cite{dumitriu03th} as Theorem 8.5.3 (the proof is virtually the same as in \cite{dumitriu03th} and we will  not repeat it here). The second one is a rewrite of a particular case ($z_a=0$ and formula (4.14a)) of formula (4.36b) in \cite{Forrester_poly}.

\begin{lemma} \label{duality_principleH}
Let $\kappa'$ denote the conjugate partition to $\kappa$ (obtained by transposing the rows and columns in the Young tableau). Then the following is true:
\begin{eqnarray*}
\mathcal{E}_{\alpha}^{H} \left [ \frac{J_{\kappa}^{\alpha}(x_1, \ldots, x_n)}{J_{\kappa}^{\alpha} (I_n)} \right ] = (-\alpha)^{-k/2} \mathcal{E}_{1/\alpha}^{H} \left [ \frac{J_{\kappa'}^{1/\alpha}(y_1, \ldots, y_m)}{J_{\kappa'}^{1/\alpha}(I_m)} \right ]~.
\end{eqnarray*}
\end{lemma}

\begin{lemma} \label{duality principleL}
The following identity is true:
\begin{eqnarray*}
\mathcal{E}_{\alpha, a}^{L} \left [ \frac{J_{\kappa}^{\alpha}(x_1, \ldots, x_n)}{J_{\kappa}^{\alpha} (I_n)} \right ] = \prod_{(x,y) \in \kappa} \left ( a - \frac{x}{\alpha} + y \right )~.
\end{eqnarray*}
In particular, for $a = \frac{n}{\alpha \gamma}$, 
\begin{eqnarray*}
\mathcal{E}_{\alpha, \frac{n}{\gamma \alpha}}^{L} \left [ \frac{J_{\kappa}^{\alpha}(x_1, \ldots, x_n)}{J_{\kappa}^{\alpha} (I_n)} \right ] = \prod_{(x,y) \in \kappa} \left ( \frac{n}{\alpha \gamma} - \frac{x}{\alpha} + y \right )~.
\end{eqnarray*}
\end{lemma}

We can now prove the two main results of this section.

\begin{theorem} \label{palindromeH}
For $k$ an even integer, let
\begin{eqnarray*}
\mathcal{E}_{\alpha}^{H} \left [ p_{[k]}(x_1, \ldots, x_n) \right ] = \sum_{j=0}^{\frac{k}{2}} f(\alpha, j) ~ n^{\frac{k}{2} +1-j}~,
\end{eqnarray*}
with $p_{[k]}$ being the power-sum corresponding to partition $[k]$.
Then $f(\alpha, j)$ is an integer-coefficient polynomial in $1/\alpha$ of degree at most $k/2$ such that 
\[
f(\alpha, j) = (-\alpha)^{-k+j} f(1/\alpha, j)~. 
\]
\end{theorem}

%\begin{corollary}  \label{perturb_H}
%If $H_{\beta, n}$ denotes a matrix from the scaled $\beta$-Hermite ensemble \ref{} (recall $\beta = 2/\alpha$), then  
%\[
%\lim_{n \rightarrow \infty} E_{\beta}^{H} \left [ \mbox{tr}(H_{\beta, n}^k) \right] - \]
%\end{corollary}
\begin{remark}
When we scale the ensembles, we have to multiply the expectation by $ (2 n \beta)^{-k/2} = \left(\frac{\alpha}{4n} \right)^{k/2}$, which means that 
\begin{eqnarray*}
\frac{1}{n} E_{\alpha}^{H} \left [ p_{[k]}(x_1, \ldots, x_n) \right ] = \frac{1}{4^{k/2}} \sum_{j=0}^{\frac{k}{2}} \alpha^{k/2} f(\alpha, j) ~ n^{-j}~.
\end{eqnarray*}
\end{remark}

The corollary below follows.

\begin{corollary} \label{linearH} It follows that $g(\alpha, j):=\alpha^{k/2} f(\alpha, j)$ is an integer-coefficient polynomial of $\alpha$ for which 
\[
g(\alpha, j) = (- \alpha)^j g(\frac{1}{\alpha}, j)~,
\]
so the degree of $g(\alpha, j)$ is at most $j$. This yields, in particular, $g(\alpha, 1) = c_k (\alpha - 1)$, with $c_k$ a constant depending on $k$. 
\end{corollary}

Similarly, for the Laguerre ensembles, we have the following theorem.

\begin{theorem} \label{palindromeL}
Let $ a= \frac{n}{\alpha \gamma}$, and 
\begin{eqnarray*}
\mathcal{E}_{\alpha, a}^{L} \left [ p_{[k]}(x_1, \ldots, x_n) \right ] = \sum_{j=1}^{k+1} ~\sum_{r=0}^{j-1}  ~f(\alpha, j, r) ~ \frac{n^j}{\gamma^r}~,
\end{eqnarray*}
with $p_{[k]}$ being the power-sum corresponding to partition $[k]$.
Then $f(\alpha, j, r)$ is a polynomial in $1/\alpha$ of degree at most $k$ such that
\[
f(\alpha, j, r) = (-\alpha)^{-k-j+1} f(1/\alpha, j, r)~.
\]
\end{theorem}

\begin{remark}
When we scale the ensembles, we have to multiply the expectation by $ \left(\frac{\gamma}{n \beta} \right)^{-k} = \left(\frac{\gamma \alpha}{2 n} \right)^{k}$, which means that 
\begin{eqnarray*}
\frac{1}{n} E_{\alpha, a}^{L} \left [ p_{[k]}(x_1, \ldots, x_n) \right ] = \frac{1}{2^k} \sum_{j=1}^{k+1} ~\sum_{r=0}^{j-1}  ~\alpha^{k} f(\alpha, j, r) ~ n^{-k+j-1} ~\gamma^{k-r}~.
\end{eqnarray*}
\end{remark}

The corollary below follows.

\begin{corollary} \label{linearL} It follows that $g(\alpha, j, r) = \alpha^{k} f(\alpha, j, r)$ is an integer-coefficient polynomial of $\alpha$ for which
\[
g(\alpha, j, r)= (-\alpha)^{k-j+1} g(1/\alpha, j,r)~, 
\]
so that the degree of $g(\alpha, j, r)$ is at most $k-j+1$. This yields, in particular, $g(\alpha, k, r) = c_{k, r} (\alpha- 1)$.
\end{corollary}

\vspace{.5cm}

\noindent \textit{Proof of Theorem \ref{palindromeH}.} Note that $1/\alpha = 2/\beta$ and that $p_{[k]}(x_1,\ldots,x_n) =$tr$(X^k)$, for any matrix $X$ with eigenvalues $x_1, \ldots, x_n$. By using the unscaled matrix model $H_{\beta, n}$ for the $\beta$-Hermite ensembles found in Table \ref{tabelu2}, one can obtain, as in Application 3 of \cite{dumitriu02} (more precisely, from Corollary 4.3), that $\mathcal{E}_{\alpha}^H [p_{[k]}(x_1, \ldots, x_n)]$ is an integer-coefficient polynomial in $1/\alpha$ of degree $k/2$, and a polynomial in $n$ of degree $k/2+1$. Hence $f(\alpha, j)$ is an integer-coefficient polynomial in $1/\alpha$ of degree at most $k/2$.

Let us now express $p_{[k]}$ in Jack polynomial basis:
\begin{eqnarray} \label{p_into_jacks}
p_{[k]} = \sum_{\lambda \vdash k} c_{\lambda}(\alpha) ~J_{\lambda}^{\alpha}~,
\end{eqnarray}
omitting the variables for simplicity.

Let $\mathbb{Q}(\alpha)$ be the field of all rational functions of $\alpha$ with rational coefficients.

Let $\Lambda \times \mathbb{Q}(\alpha)$ be the vector space of all symmetric polynomials of bounded degree with coefficients in $\mathbb{Q}(\alpha)$.

For every $0 \neq \theta \in \mathbb{Q}(\alpha)$, define the $\mathbb{Q}(\alpha)$-algebra automorphism $\omega_{\theta} : \Lambda \times \mathbb{Q}(\alpha) \rightarrow \Lambda \times \mathbb{Q}(\alpha)$ by the condition $\omega_{\theta}(p_k) = (-1)^{k-1} \theta p_k$, for all $k \geq 1$. This family of automorphisms appears in \cite[Chapter 10]{MacDonald_book}, and similarly in \cite{Stanley_jacks}. In particular, $\omega = \omega_1$ is known as the Macdonald\index{Macdonald, I.} involution (and can be found in \cite[Chapter 1]{MacDonald_book}).

We will use the following formula due to Stanley \cite{Stanley_jacks} which can also be found as formula (10.24) in \cite{MacDonald_book}:
\begin{eqnarray}
\label{omega_3} \omega_{\alpha} J_{\kappa}^{\alpha} & = & \alpha^{|\kappa|} ~J_{\kappa'}^{1/\alpha}~ ,
\end{eqnarray}
where $\lambda'$ is the conjugate partition of $\lambda$ (obtained from $\lambda$ by transposing rows and columns in the Young tableau).

The first step of the proof is given by the following lemma.

\begin{lemma} \label{p_into_jacks_palindrome} The coefficients in equation \eqref{p_into_jacks} satisfy
\[
c_{\lambda}(\alpha) = (-\alpha)^{1-k} c_{\lambda'}(1/\alpha)~.
\]
\end{lemma}

\begin{proof} We apply $\omega_{\alpha}$ to both sides of \eqref{p_into_jacks}, and use the fact that $\omega_{\alpha}$ is linear, together with \eqref{omega_3}, to obtain that, on one hand,
\begin{eqnarray*}
\omega_{\alpha} p_{[k]} & = & (-1)^{k-1} \alpha p_{[k]} = (-1)^{k-1} \alpha \sum_{\lambda \vdash k} c_{\lambda}(1/\alpha) J_{\lambda}^{1/\alpha}~,
\end{eqnarray*}
and
\begin{eqnarray*}
\omega_{\alpha} p_{[k]} & = & \sum_{\lambda \vdash k} c_{\lambda}(\alpha) \alpha^k J_{\lambda'}^{1/\alpha}~,
\end{eqnarray*}
on the other hand.

Since there is a unique way of writing $\omega_{\alpha} p_{[k]}$ in Jack polynomial basis, it follows that 
\[
c_{\lambda}(\alpha)  = (-\alpha)^{1-k} c_{\lambda'}(1/\alpha)~.
\]
\end{proof}

\begin{remark} Note that Lemma \ref{p_into_jacks_palindrome} does not say anything about expectations.
\end{remark}

We now write the expectation of $p_{[k]}$ over the unscaled Hermite ensemble using \eqref{p_into_jacks}:
\begin{eqnarray*}
\mathcal{E}_{\alpha}^{H} \left [ p_{[k]} (x_1, \ldots, x_n) \right ]  = \sum_{\lambda \vdash k} c_{\lambda}(\alpha) ~J_{\lambda}^{\alpha} (I_n) ~ \mathcal{E}_{\alpha}^{H} \left [ \frac{J_{\lambda}^{\alpha}(x_1, \ldots, x_n)}{J_{\lambda}^{\alpha}(I_n)} \right ]~.
\end{eqnarray*}

We know (for example from \cite[page]{Stanley_jacks}) that $J_{\lambda}^{\alpha}(I_n) = \prod_{(x,y) \in \lambda} (m - x + \alpha y)$; hence
\begin{eqnarray} \label{unu}
\mathcal{E}_{\alpha}^{H} \left [ p_{[k]} (x_1, \ldots, x_n) \right ]  = \sum_{\lambda \vdash k} c_{\lambda}(\alpha) ~ \mathcal{E}_{\alpha}^{H} \left [ \frac{J_{\lambda}^{\alpha}(x_1, \ldots, x_n)}{J_{\lambda}^{\alpha}(I_n)} \right ] ~\prod_{(x,y) \in \lambda} (n - x + \alpha y)~.
\end{eqnarray}

By Lemma \ref{duality_principleH}, $\mathcal{E}_{\alpha}^{H} \left [ \frac{J_{\lambda}^{\alpha}(x_1, \ldots, x_n)}{J_{\lambda}^{\alpha}(I_n)} \right ]$ does \emph{not} depend on $n$. Write
\[
\prod_{(x,y) \in \lambda} (n - x + \alpha y) = \sum_{j=0}^{|\lambda|} b_{\lambda}(j, \alpha) n^{j}~;
\]
we have
\[
\prod_{(x,y) \in \lambda} (n - x + \alpha y) = \prod_{(y, x) \in \lambda'} (n + \alpha (y - \frac{x}{\alpha}))~,
\]
and consequently
\begin{eqnarray} \label{un}
b_{\lambda}(j, \alpha) = (-\alpha)^{k-j} b_{\lambda'}(j, 1/\alpha)~.
\end{eqnarray}

Using the above, \eqref{unu}, Lemma \ref{p_into_jacks_palindrome}, Lemma \ref{duality_principleH},  and substituting $\frac{k}{2}+1-j$ for $j$ (in the power index of $n$), we obtain the statement of Theorem \ref{palindromeH}.
\qed

\vspace{.25cm}

\noindent \textit{Proof of Theorem \ref{palindromeL}.} The fact that $f(\alpha, j, r)$ is a polynomial in $1/\alpha$ of degree $k$ follows similarly to Corollary 4.3 in Application 3 of \cite{dumitriu02}. 

We write
\begin{eqnarray}
\mathcal{E}_{\alpha, a}^{L} \left [ p_{[k]} (x_1, \ldots, x_n) \right ]  & = & \sum_{\lambda \vdash k} c_{\lambda}(\alpha) ~ \mathcal{E}_{\alpha, a}^{L} \left [ \frac{J_{\lambda}^{\alpha}(x_1, \ldots, x_n)}{J_{\lambda}^{\alpha}(I_n)} \right ] ~\prod_{(x,y) \in \lambda} (n - x + \alpha y)~, \\
\label{doi}& = & \sum_{\lambda \vdash k} c_{\lambda}(\alpha) \prod_{(x,y) \in \lambda} \left( \frac{n}{\alpha \gamma} - \frac{x}{\alpha} + y \right ) ~\prod_{(x,y) \in \lambda} (n - x + \alpha y)~.
\end{eqnarray}

If we write
\[
\prod_{(x,y) \in \lambda} \left( \frac{n}{\alpha \gamma} - \frac{x}{\alpha} + y \right ) = \sum_{j=0}^{|\lambda|} \tilde{b}_{\lambda}(j, \alpha) n^{j} \gamma^{-j}~,
\]
it is not hard to see that 
\begin{eqnarray} \label{deux}
\tilde{b}_{\lambda}(j, \alpha) = (-\alpha)^{-k-j} ~\tilde{b}_{\lambda'}(j, 1/\alpha)~.
\end{eqnarray} 

Using \eqref{un}, \eqref{deux}, and Lemma \ref{p_into_jacks_palindrome}, we obtain the statement of Theorem \ref{palindromeL}.
\qed

\subsection{Computing the $\beta$-independent part of the deviation}
%\subsection{Generating functions and second-order differential equations}

In this section we will examine the deviation at $\beta = \infty$ for the Hermite and Laguerre ensembles. In proving this we employ a simple differential equations trick that will allow us to compute the zero and first order terms in the mean of the eigenvalue distributions at $\beta = \infty$.

Given a function $y(x)$ which satisfies the second order homogeneous differential equation
\[
f(x) y''(x) +g(x) y'(x) +h(x) y(x) = 0~,
\]
denote by $m(x)$ the function $m(x) = y'(x)/y(x)$ (with poles at the zeroes of $y(x)$).

\begin{proposition} \label{gf}
The function $m(x)$ satisfies the first-order differential algebraic equation
\[
m^2(x) + \frac{g(x)}{f(x)} m(x) + \frac{h(x)}{f(x)} + m'(x) = 0~.
\]
\end{proposition}

The proof is immediate. 

\begin{remark}
If the function $y(x)$ is a polynomial with a finite number $k$ of distinct roots, $m(x)$ is the generating function for the powers of $y$'s roots. 
\end{remark}

%\section{First-order fluctuations: the mean}

%In this section we compute the mean of the fluctuation process, as given in Theorem \ref{}. 

Let $\tilde{H}_{\beta,n}$ and $\tilde{L}_{\beta, n}^{a}$ denote matrices from the $\beta$-Hermite, respectively, $\beta$-Laguerre ensembles. As in \cite{dumitriu03th}, to obtain the deviation, we once again will examine the averaged traces of powers of the matrices $\tilde{H}_{\beta,n}$ and $\tilde{L}_{\beta, n}^{a}$, this time looking at the first-order terms. 

As a consequence of Corollaries \ref{linearH} and \ref{linearL}, for $k$ an even positive integer in the Hermite case, and $k$ an arbitrary positive integer in the Laguerre case with parameter $a =(n \beta)/(2 \gamma)$,
\[
\frac{1}{n} E_{\beta}^{H} \left[ p_{[k]}(x_1, \ldots, x_n) \right] = c_k + c_{k}^{1} \frac{\frac{2}{\beta}-1}{n} + O(n^{-2})~,
\]
while 
\[
\frac{1}{n}E_{\beta}^{L,a} \left[ p_{[k]}(x_1, \ldots, x_n) \right] = c_k(\gamma) + c_{k}^{1}(\gamma) \frac{\frac{2}{\beta}-1}{n} + O(n^{-2})~.
\]

It follows that the deviation is given by the moments $c_{k}^{1}$, respectively, $c_{k}^{1}(\gamma)$, times the scaling factor $\frac{2}{\beta}-1$. Equivalently, if we examine the generating functions 
\begin{eqnarray*}
m(n, \beta, x) & = & \frac{1}{n} E_{\beta}^{H} \left[ \sum_{i=1}^n \frac{1}{x - \lambda_i} \right] \\
& = & \frac{1}{n} \sum_{k=0}^{\infty} \frac{E_{\beta}^{H} \left[ p_{[k]}(x_1, \ldots, x_n) \right]}{x^{k+1}} \\
& = & \sum_{k=0}^{\infty} \left ( \frac{c_k}{x^{k+1}}+ \frac{1}{n} \left(\frac{2}{\beta} -1 \right)\frac{c_{k}^{1}}{x^{k+1}} + \ldots \right)
\end{eqnarray*}
and 
\begin{eqnarray*}
m(n, a, \beta, x) & = & \frac{1}{n} E_{\beta, a}^{L} \left[ \sum_{i=1}^n \frac{1}{x - \lambda_i} \right]  \\
& = & \frac{1}{n} \sum_{k=0}^{\infty} \frac{E_{\beta, a}^{L} \left[ p_{[k]}(x_1, \ldots, x_n) \right]}{x^{k+1}} \\
& = & \sum_{k=0}^{\infty} \left ( \frac{c_k}{x^{k+1}} + \frac{1}{n} \left( \frac{2}{\beta} -1 \right)\frac{c_{k}^{1}(\gamma)}{x^{k+1}} + \ldots \right)
\end{eqnarray*}
then in order to find the first order asymptotics of $m(n, \beta, x)$ and $m(n,a, \beta, x)$, it is enough to compute $\sum\limits_{k=0}^{\infty} \frac{c_k^{1}}{x^{k+1}}$, respectively, $\sum\limits_{k=0}^{\infty} \frac{c_{k}^{1}(\gamma)}{x^{k+1}}$.

We will do this by keeping $n$ fixed, letting $\beta \rightarrow \infty$, and computing the zero and first order terms in $\frac{1}{n}$ in the generating function of the resulting matrix ensemble.

\begin{remark} 
Knowing that the quantities $\frac{1}{n} E_{\beta}^{H} \left[ p_{[k]}(x_1, \ldots, x_n) \right]$ and $\frac{1}{n}E_{\beta}^{L,a} \left[ p_{[k]}(x_1, \ldots, x_n) \right]$ are polynomial expressions in $\frac{1}{n}$ and $\frac{2}{\beta}$ (which follows immediately by rescaling the results of  Theorems \ref{palindromeH} and \ref{palindromeL}) is the key factor in allowing us to let $\beta \rightarrow \infty$ while keeping $n$ fixed. \end{remark}

To aid us in our calculations, we will make use of a beautiful property of the $\chi_{r}$ distribution, which will allow us to replace our tridiagonal models with even simpler ones. We give this property as a Proposition.

\begin{proposition} \label{chi_simple} Let $\{X_n\}_{n \in \mathbb{N}}$ be a set of random variables with distributions $\chi_{r_n}$, $n \in \mathbb{N}$, such that $r_n \rightarrow \infty$ as $n \rightarrow \infty$. Then the sequence $\{X_n - \sqrt{r_n}\}_{n \in \mathbb{N}}$ converges in distributions to a centered normal of variance $1/2$.
\end{proposition}

It follows then that, as we have already shown in \cite{dumitriu04a}, that given $n$ and $\gamma \leq 1$ fixed, the entries of the scaled random matrices $\tilde{H}_{\beta, n}$ and $\tilde{L}_{\beta, n}^{a} = B_{\beta, n}^{a} (B_{\beta, n}^{a})^{T}$ (with $a =(n \beta)/(2\gamma)$) converge as $\beta \rightarrow \infty$ to the following models:
\begin{eqnarray} \label{fixed_h_matrix}
\tilde{H}_{\beta, n} \rightarrow H = \frac{1}{2\sqrt{n}} \left( \begin{array}{cccccc} 0 & \sqrt{n-1} & &&  & \\
				\sqrt{n-1} & 0 & \sqrt{n-2} &  & &\\
				           & \sqrt{n-2} & 0 &  & &\\
					   & & &\ddots & &\\
					%   & & &  0 & \sqrt{2} &\\
					   & & & &0 & \sqrt{1} \\
					   &  && &\sqrt{1} & 0 \end{array} \right )~.
\end{eqnarray} 
respectively to 
\footnotesize{
\begin{eqnarray} \label{fixed_l_matrix}
\tilde{L}_{\beta, n}^{a} \rightarrow L_{\gamma} = \frac{\gamma}{n} \left ( \begin{array}{ccccc} \frac{n}{\gamma} & \sqrt{\frac{n}{\gamma}} ~\sqrt{n-1} & & &\\
					  \sqrt{\frac{n}{\gamma}} ~\sqrt{n-1} & \frac{n}{\gamma}+n-2  & \sqrt{\frac{n}{\gamma}-1} ~\sqrt{n-2} & &\\
& \sqrt{\frac{n}{\gamma}-1} ~\sqrt{n-2} &  \frac{n}{\gamma}+n -3 & & \\
& & \ddots & & \\
& & & \sqrt{\frac{n}{\gamma}-n+2} ~\sqrt{2} &\\
& & \sqrt{\frac{n}{\gamma}-n+2} ~\sqrt{2} & \frac{n}{\gamma}-n+4 & \sqrt{\frac{n}{\gamma}-n+1}~ \sqrt{1} \\
& & & \sqrt{\frac{n}{\gamma}-n+1} ~\sqrt{1} & \frac{n}{\gamma}-n+2  
\end{array} \right )
\end{eqnarray}}
\normalsize

Note also that if we denote by $\tilde{B}_{\beta, n}^{a}$ the matrix such that $ \tilde{L}_{\beta, n}^{a} = \tilde{B}_{\beta, n}^{a} (\tilde{B}_{\beta, n}^{a})^{T}$, then 
\begin{eqnarray} \label{fixed_b_matrix}
\tilde{B}_{\beta,n}^{a} \rightarrow B_{\gamma} = \sqrt{\frac{\gamma}{n}} \left ( \begin{array}{ccccc} \sqrt{\frac{n}{\gamma}} & & & & \\ \sqrt{n-1} & \sqrt{\frac{n}{\gamma} -1} & & & \\ & \sqrt{n-2} & \sqrt{\frac{n}{\gamma} -2} & & \\ & & \ddots && \\ & & \sqrt{2} & \sqrt{\frac{n}{\gamma} - n+2} & \\ & & & \sqrt{1} & \sqrt{\frac{n}{\gamma} -n+1} 
\end{array} \right )
\end{eqnarray}
and that $L_{\gamma} = B_{\gamma} B_{\gamma}^T$.

Since $H$, $L_{\gamma}$, and $b_{\gamma}$ are \emph{\textbf{non-random}} matrices, all expectations are exact. 

\subsubsection{$\beta = \infty$, Hermite case} \label{binfh}

The matrix $H$ (see \eqref{fixed_h_matrix}) has as eigenvalues $h_1/\sqrt{2n}, \ldots, h_n/\sqrt{2n}$, where $h_1, \ldots, h_n$ are the roots of the $n$th Hermite polynomial $H_n(x)$ (this can be easily deduced from the three-term recurrence for the Hermite polynomials, see for example \cite{mathworld_herm}). For a more detailed description of the properties of this matrix, see \cite{dumitriu04a}.

It follows that the generating function we need to compute is 
\[
\tilde{m}(n, x) = \frac{1}{n} \sum_{i=1}^n \frac{1}{x - \frac{h_i}{\sqrt{2n}}}~.
\]

We use the well-known identity
\begin{eqnarray} \label{charpoly}
\sum_{i=1}^{n} \frac{1}{x - x_i} = \frac{p'(x)}{p(x)}~, 
\end{eqnarray}
where $x_i$ are distinct values, and $p(x)$ is the polynomial whose roots the $x_i$s are, to obtain that
\[
\tilde{m}(n,x) = \sqrt{\frac{n}{2}} ~\frac{H_{n}'(x \sqrt{2n})}{ H_{n}(x \sqrt{2n})}~,
\]
and by applying Proposition \ref{gf}, we get that $\tilde{m}(n,x)$ satisfies the differential equation
\begin{eqnarray} \label{herm_gf}
\left(\tilde{m}(n,x)\right)^2 - 4x\tilde{m}(n,x) + 4 + \frac{\tilde{m}'(n,x)}{n} = 0~.
\end{eqnarray}

Writing $\tilde{m}(n, x) = m_0(x) - \frac{1}{n} m_1(x) + O(n^{-2})$, we obtain from \eqref{herm_gf} that 
\begin{eqnarray*}
m_0(x) & = & 2\left(x - \sqrt{x^2-1} \right) \\
m_1(x) & = & \frac{\left(x-\sqrt{x^2-1} \right)}{2(x^2-1)}~.
\end{eqnarray*}

Computing the inverse Cauchy transform for $m_0(x)$ and $m_1(x)$ yields the semicircle distribution and, respectively 
\begin{eqnarray} \label{fluct_mean_H}
\mu_H(x) = \left \{ \begin{array}{cl} \frac{1}{4} (\delta_1(x) + \delta_{-1}(x)) - \frac{1}{2\pi} \frac{1}{\sqrt{1 - x^2}}, & \mbox{if $x \in [-1, 1]$}~, \\ 0~, & \mbox{otherwise.} \end{array} 
\right .
\end{eqnarray}

We have thus proved the following result.

\begin{lemma} \label{mean_H}
For any polynomial $P$,   
\[
E_{\beta}^H \left [\sum_{i=1}^n P(x_i) \right ] - n~\int_{-1}^{1} P(x) s(x) dx \longrightarrow \left(\frac{2}{\beta} -1\right) \int_{-1}^{1} P(x) \mu_H(x) dx~,
\]
as $n \rightarrow \infty$.
\end{lemma}

\begin{remark} As a side note, in the computation above we have provided yet another way to obtain the semicircle law for all $\beta$. \end{remark}

\subsubsection{$\beta = \infty$, Laguerre case} \label{binfl}

The matrix $L_{\gamma}$ has as eigenvalues $\gamma l_1/n, \ldots, \gamma l_n/n$, where $l_1, \ldots, l_n$ are the roots of the $n$th Laguerre polynomial $L_n^{n(1/\gamma - 1)}(x)$ (see \cite{mathworld_lag}) (this can be easily deduced from one of the many recurrences for Laguerre polynomials, found for example as $(26)$ in \cite{mathworld_lag}). To get a more detailed description of the properties of this matrix, refer to \cite{dumitriu04a}, substituting $n(1/\gamma)-1$ for $\gamma$.

Since we need to rescale the eigenvalues by an additional $n$, it follows that the quantity of interest is 
\[
\tilde{m}(n, \gamma, x) = \frac{1}{n} \sum_{i=1}^n \frac{1}{x - \frac{\gamma l_i}{n}}~,
\]

Once again we use identity \ref{charpoly} and obtain that 
\[
\tilde{m}(n,\gamma, x) =  ~\frac{1}{n}~\frac{\left(L_{n}^{n(1/\gamma-1)}\right)' \left (\frac{x n}{\gamma} \right )}{ L_{n}^{n(1/\gamma-1)}\left(\frac{xn}{\gamma} \right )}~,
\]
and by applying Proposition \ref{gf}, we get that $\tilde{m}(n,\gamma, x)$ satisfies the algebraic differential equation
\begin{eqnarray} \label{lag_gf}
\gamma \left(\tilde{m}(n,\gamma, x)\right)^2 - \tilde{m}(n,\gamma,x)\left(1 - \frac{1}{x} +\frac{\gamma}{x} \right)  + \frac{1}{x}  + \frac{\gamma \tilde{m}'(n,\gamma, x)+\frac{\gamma \tilde{m}(n, \gamma, x)}{x}}{n} = 0~.
\end{eqnarray}

Writing $\tilde{m}(n, \gamma, x) = m_0(x) - \frac{1}{n} m_1(x) + O(n^{-2})$, we obtain from \eqref{lag_gf} that 
\begin{eqnarray*}
m_0(x) & = & \frac{x+ \gamma - 1 - \sqrt{ (x - (\sqrt{\gamma}-1)^2)(x - (\sqrt{\gamma}+1)^2)}}{2 \gamma x}~, \\
m_1(x) & = &  \frac{x - \gamma - 1 -\sqrt{ (x - (\sqrt{\gamma}-1)^2)(x - (\sqrt{\gamma}+1)^2)}}{2(x - (\sqrt{\gamma}-1)^2)(x - (\sqrt{\gamma}+1)^2)}~.
\end{eqnarray*}

By calculating the inverse Cauchy transforms of $m_0(x)$ and $m_1(x)$, one obtains the Mar\v{c}enko-Pastur distribution, respectively 
\begin{eqnarray}
t_{\gamma}(x) = \left \{ \begin{array}{cl} \frac{1}{4} \delta_{b}(x) + \frac{1}{4} \delta_a(x) - \frac{1}{2 \pi} \frac{1}{\sqrt{(x-a)(b-x)}}~, & \mbox{if $x \in [a, b]$}~, \\ 0~, & \mbox{otherwise~,} \end{array} 
\right .
\end{eqnarray}
with $a = (\sqrt{\gamma}-1)^2$, $b = (\sqrt{\gamma}+1)^2$.

We have thus proved the following result.

\begin{lemma} \label{mean_L}
For any polynomial $P$,   
\[
E_{\beta, a}^L \left [\sum_{i=1}^n P(x_i) \right ] - n~\int_{a}^{b} P(x) e_{\gamma}(x) dx \longrightarrow \left(\frac{2}{\beta} -1\right) \int_{a}^{b} P(x) t_{\gamma}(x) dx~,
\]
as $n \rightarrow \infty$.
\end{lemma}

\section{Fluctuation of the semicircle and Mar\v{c}enko-Pastur laws} \label{gauss}

In this section we compute the fluctuations terms for the $\beta$-Hermite and $\beta$-Laguerre ensembles; we show that the fluctuation of the trace of any given power of the matrix corresponding to the ensemble tends to a Gaussian.

The essence of the argument is simple. We will think of the random matrix as the sum between the non-random matrix of means (which we can also think about roughly as the $\beta = \infty$ non-random matrix), and a random matrix of the centered entries, and do some obvious computations of traces of powers. Much of the work goes into the technical carefulness to provide a complete argument, but the reader should not let this detract them from the simplicity of the idea, which is based on formula which gives, for a matrix $T$,
\[
\mbox{tr}(T^k) = \sum_{i_1, \ldots, i_k} t_{i_1 i_2} t_{i_2 i_3} \ldots t_{i_k i_1}~,
\]
with the proviso that the above sum is especially simple when the matrix $T$ is tridiagonal. 

We provide here a heuristic explanation for the Hermite case, just for the purpose of emphasizing that the main idea is simple. 

One can think of the random matrix, for all practical purposes, as 
\[
T \approx T_{\beta = \infty} + \frac{1}{\sqrt{\beta n}} G~,
\]
where $G$ is symmetric tridiagonal matrix of $O(1)$-variance Gaussians (similar to the decomposition we used in \cite{dumitriu04a}), with entries which are mutually independent, up to the symmetry condition. 

Then for any polynomial $h$,
\[
\mbox{tr}(h(T)) \approx \mbox{tr}(h(T_{\beta = \infty})) + \frac{1}{\sqrt{\beta n}} \mbox{tr}(h'(T_{\beta = \infty}) G);
\]
the second term is clearly normally distributed, and all we have to do is compute the variance and show it is finite (which we can achieve by examining the cases when $h$ is a monomial).

It is worth noting that, in principle, one should be able to do this computation for continuously differentiable functions $h$ with some additional conditions imposed by the fact that the variance needs to be finite.

The technicalities arise because $n \rightarrow \infty$ and the equalities above are just approximations, but this should not detract from the main idea. As we will see, using the method of moments will show that we do not need $G$'s entries to be Gaussian (or even \emph{approximately} Gaussian) in order for the \emph{fluctuation} to monomials $h$ to be Gaussian.

\subsection{The $\beta$-Hermite case} \label{gaus_herm}

We write the scaled matrix $\tilde{H}_{\beta,n}$ as
\begin{eqnarray} \label{sc_herm}
\tilde{H}_{\beta, n} = A + \frac{1}{\sqrt{n \beta}} Y~,
\end{eqnarray}
where $A = E_{\beta}^{H} [ \tilde{H}_{\beta, n}]$ is the symmetric tridiagonal matrix of mean entries ($A(i, i+1) = \frac{1}{2\sqrt{n \beta}}E[\chi_{(n-i)\beta}]$; all other entries are $0$), and $Y$ is symmetric tridiagonal matrix of centered variables (with a diagonal of independent Gaussians of variance $1/2$). Technically, $A$ and $Y$ depend both on $n$ and on $\beta$; we drop these indices from notation for the sake of simplicity.

\begin{remark} \label{convergence} From Proposition \ref{chi_simple}, we know that if $A = (a_{ij})_{\mbox{{\small $1 \leq i,j \leq n$}}}$, then for any $\epsilon>0$, there exists $i_{\epsilon} \in \mathbb{N}$ such  that 
\[
\left (2\sqrt{n \beta} ~a_{i,i+1} - \sqrt{(n-i) \beta} \right)  < \epsilon ~~\mbox{for any} ~~ i \leq (n-i_{\epsilon})~~,
\] and that $Y_{i, i+1} \rightarrow N(0, 1/8)$ in distribution as $i \rightarrow \infty$, while $Y_{i,i} \sim N(0, 1/2)$ for all $i$. \end{remark}

%We restate this as follows:
\begin{remark}\label{boundedness}\textbf{(bounded moments)}  Note that the entries of $A$ are \textbf{\emph{bounded}}, both from below and from above; we will think of them as $O(1)$. Similarly, for any $k$ and $l$ finite, we know from the above that there exists an $M$ such that 
\[
E[\prod_{i=1}^{kl} Y_{j_i, j'_i}^{c_i}] \leq M, 
\]
for all $0 \leq c_i \leq kl$, and for all $j_1, \ldots, j_{kl}$ and $j'_1, \ldots, j'_{kl}$ such that $|j_{i} - j'_{i}| \leq 1$. 
\end{remark}

Given integers $k$ and $n$, consider the random variable
\begin{eqnarray*} 
\omega_k(n) & = & \mbox{tr}(\tilde{H}_{\beta,n}^k) - E^{H}_{\beta}[\mbox{tr}(\tilde{H}_{\beta,n}^k)]~.
% \\
%\eta_k(n) & =& \mbox{tr}(M_h^k) ~-E[\mbox{tr}(M_h^k)]~.
\end{eqnarray*}

%\begin{itemize} \item[\textit{Claim 1.}] 
%$\eta_k(n) \rightarrow N(0, \sigma_k^2) ~\mbox{as}~ n \rightarrow \infty~,$
%where $\sigma_k^2 = ????$,
%\item[\textit{Claim 2.}] 
\begin{Claim} \label{claim1} For any fixed integers $k$ and $l$,
\[\lim_{n \rightarrow \infty} E[(\omega_k(n))^l] = ~\left \{\begin{array}{ll} 
 \left(\frac{2}{\beta}\right)^{l/2} \frac{1}{2^{(k+1)l}} ~k^{l/2}~ {k \choose \frac{k}{2}}^l, &~~~\mbox{if $k$, $l$ are even,} \\
 \left(\frac{2}{\beta}\right)^{l/2} \frac{1}{2^{kl}} ~k^{l/2} ~{k-1 \choose \frac{k-1}{2}}^l, &~~~\mbox{if $k$ is odd and $l$ is even}~,\\ ~~0, &~~~\mbox{if $l$ is odd.} \end{array} \right .
\]
\end{Claim}
\begin{Claim} \label{claim2} For any fixed integers $k_1$ and $k_2$, 
\[\lim_{n \rightarrow \infty} \mbox{Cov}(\omega_{k_1}(n), \omega_{k_2}(n)) = \left \{ \begin{array}{ll} \frac{2}{\beta} ~\frac{1}{2^{k_1+k_2}}~ \frac{2k_1k_2}{k_1+k_2} {k_1-1 \choose \frac{k_1-1}{2}} {k_2 \choose \frac{k_2-1}{2}}~, & \mbox{if $k_1, k_2$ are odd,} \\ 
\frac{2}{\beta} ~\frac{1}{2^{k_1+k_2+2}}~ \frac{2k_1k_2}{k_1+k_2} {k_1 \choose \frac{k_1}{2}} {k_2 \choose \frac{k_2}{2}}~, & \mbox{if $k_1, k_2$ are even},\\~~0~,& \mbox{otherwise}~.\end{array} \right .\]
\end{Claim}

\begin{remark} \label{red_rem}
Claims \ref{claim1} and \ref{claim2} show that the centered fluctuation of the $\beta$-Hermite ensembles describes the Gaussian process on monomials defined in Theorem \ref{main_herm}.
\end{remark}

For the remainder of this section, we will prove Claims \ref{claim1} and \ref{claim2}. To give the reader a rough idea of where the calculations will lead, we provide below an intuition of what we will be doing.

\vspace{.25cm}

\noindent \textit{\textbf{Intuitive explanation}}. The first step is to note that $tr(T^k)$ is a sum of products of $k$ entries of $T$; for a tridiagonal matrix $T = (t_{i,j})_{1 \leq i, j \leq n}$ with $t_{i,j} = 0$ if $|i-j|>1$,
\[
\mbox{tr}(T^k) = \sum_{1 \leq i_1, \ldots, i_k \leq n} t_{i_1, i_2} t_{i_2, i_3} \ldots t_{i_k, i_1}~,
\]
where the sum needs to be taken only over the sequences $i_1, \ldots, i_k$ such that $|i_{j} - i_{j+1}| \leq 1$, for all $j=1, \ldots, k-1$, and also $|i_k - i_1| \leq 1$. 

We have a sliding ``window'' of size $k$ down the diagonal of the matrix $T$ in which we take products of powers of the elements. In particular, for the matrix $A$, this is easy to visualize, because with the exception of a finite bottom right corner, the entries of $A$ in any finite window look roughly the same. 

The second step is to identify the significant terms, i.e. the terms that have non-zero asymptotical contributions. Roughly speaking, these will be the terms which will contain \emph{precisely} one element of $Y$, and all the others from $A$. Nothing surprising here, as $Y$ is scaled by $\frac{1}{\sqrt{n \beta}}$ (see \eqref{sc_herm}).

Finally, we will compute the contribution from the significant terms and show it agrees with the result of Claim \ref{claim1}. Then we will note that the same reasoning yields the result of Claim \ref{claim2}.

\vspace{.5cm}

We now proceed to make the above intuitive description rigorous. We need to introduce some notation. 

\begin{definition} For  given $n$ and $p$, we denote by $\mathcal{S}_{n,p} \subset \{1, \ldots, n\}^p$ the set of sequences of integers $i_1, \ldots, i_p$ such that $(i_1, \ldots, i_p) \in \{1, \ldots, n\}^p$ and $|i_j - i_{j+1}| \leq 1$ for all $j=1, \ldots, p-1$, and also $|i_p - i_1| \leq 1$. 

We denote by $\mathcal{I}$ an element of $\mathcal{S}_{n,p}$, and we denote by \[
(T)_{\mathcal{I}} := t_{i_1, i_2} t_{i_2, i_3} \ldots t_{i_p, i_1}~,
\]
where $(i_1, i_2, \ldots, i_p) =: \mathcal{I}$. \end{definition}

For a given $\mathcal{I} \in \mathcal{S}_{n,p}$, note that we can break up the sequence $(i_1, \ldots, i_p)$ into concatenations of sequences 
\[
J = ((i_{p_0}, \ldots, i_{p_1}), (i_{p_2}, \ldots, i_{p_3}), (i_{p_4} \ldots i_{p_5}), \ldots, (i_{p_{2q}}, \ldots, i_{p_{2q+1}}))~,
\] 
and 
\[
R = ((i_{p_1}, \ldots, i_{p_2}), (i_{p_3}, \ldots, i_{p_4}), \ldots, (i_{p_{2q-1}}, \ldots, i_{p_{2q}}))~,
\] 
such that in each of the sequences $i_{p_{2k}}, i_{p_{2k}+1}, \ldots, i_{p_{2k+1}}$ (for $k = 0, \ldots, q$) which form $J$, consecutive indices differ by exactly 1, and in addition to this
\[
(i_1, \ldots, i_p) = (i_{p_0}, \ldots, i_{p_1}, i_{p_1+1}, \ldots, i_{p_2}, i_{p_2+1}, \ldots, i_{p_3}, \ldots, i_{p_{2q+1}})~.
\]

We allow for the possibility of having empty sequences $i_{p_0}, \ldots, i_{p_1}$ in the beginning and/or $i_{p_{2q}}, \ldots, i_{p_{2q+1}}$ in the end of $J$.

\begin{remark} To give an intuition for these sequences, note that any term in tr$((H_{\beta,n})^k)$ can be thought of in terms of products of entries from $A$ and entries from $Y$; the sequences $J$ and $R$ will be overlapping ``runs'' recording the former, respectively the latter.
%\end{remark}

%\begin{remark} \label{one-to-one}
Also note that, given a fixed $\mathcal{I}$,  each $J$ satisfying the requirements above has exactly one $R = R(J)$ corresponding to it, and that to different $R$s correspond different $J$s. Furthermore, since $k$ is finite, a given concatenation of sequences $R$ may corresponds only to a finite number of sequences $\mathcal{I}$ (since all indices must be within $k$ of each other!).
\end{remark}

\begin{definition} \label{j_r}
We define the set $\mathcal{J}$ as the set of pairs $(J,R)$ described above. For a tridiagonal matrix $T$, we define 
\[
(T)_{J} = t_{i_{p_0}, i_{p_0}+1} \ldots t_{i_{p_1}-1, i_{p_1}} t_{i_{p_2} i_{p_2}+1} \ldots t_{i_{p_3}-1, i_{p_3}} \ldots t_{i_{2q}, i_{2q}+1} \ldots t_{i_{2q+1}-1, i_{2q+1}}~;
\]
similarly, 
\[
(T)_{R} = t_{i_{p_1}, i_{p_1}+1} \ldots t_{i_{p_2}-1, i_{p_2}} t_{i_{p_3} i_{p_3}+1} \ldots t_{i_{p_4}-1, i_{p_4}} \ldots t_{i_{2q-1}, i_{2q-1}+1} \ldots t_{i_{2q}-1, i_{2q}}
\]
\end{definition}

%\begin{itemize}
%\item 

For any sequence $\mathcal{I} \in \mathcal{S}_{n,p}$, we can write 
\begin{eqnarray} \label{descriptive}
(\tilde{H}_{\beta, n})_{\mathcal{I}} = (A+\frac{1}{\sqrt{n\beta}} Y)_{\mathcal{I}} = \sum_{(J, R) \in \mathcal{J}} \frac{1}{(n \beta)^{P/2}} (A)_{J}~ (Y)_{R}~,
\end{eqnarray}
with $P$ being the total length of the ``runs'' in the sequence $R$ (i.e., $P = (p_2-p_1+1)+(p_4-p_3+1) +\ldots+(p_{2q}-p_{2q-1}+1)$).

We have now enough information to start the proof of Claim \ref{claim1}.

\vspace{.25cm}

\noindent \textit{Proof of Claim \ref{claim1}.} First we examine 
\[
E[(\omega_k(n))^l] = E \left [ \left (\mbox{tr}(\tilde{H}_{\beta,n}^k) - E^{H}_{\beta}[\mbox{tr}(\tilde{H}_{\beta,n}^k)] \right)^l \right]~;
\]
note that 
\[
E[(\omega_k(n))^l] = E[\sum_{\mathcal{I}_1, \ldots, \mathcal{I}_l \in \mathcal{S}_{n,k}} \prod_{j=1}^l \left ( (\tilde{H}_{\beta, n})_{\mathcal{I}_{j}} - E[(\tilde{H}_{\beta,n})_{\mathcal{I}_{j}}] \right )]~.
\]

Using \eqref{descriptive}, we write
\begin{eqnarray} \label{unu1}
\!\!\!\!\!\!E[(\omega_k(n))^l] \!\!&= &\!\!E \left[ \!\!\sum_{\mbox{{\small $\begin{array}{c} \mathcal{I}_j \in \mathcal{S}_{n,k} \\ 1 \leq j \leq l \end{array}$}}} \!\!\sum_{(J_j, R_j) \in \mathcal{J}_j} \frac{1}{(n \beta)^{P_i/2}} \prod_{j=1}^l \left ( (A)_{J_i} (Y)_{R_i} - E[(A)_{J_i} (Y)_{R_i}] \right ) \right]~.
\end{eqnarray}

Since $A$ is a non-random matrix, it follows that we can rewrite \eqref{unu1} as 
\begin{eqnarray} \label{doi2}
\!\!\!\!\!\!\!\!E[(\omega_k(n))^l] \!\!&= & \!\!\!\! \sum_{\mbox{{\small $\begin{array}{c} \mathcal{I}_j \in \mathcal{S}_{n,k} \\ 1 \leq j \leq l \end{array}$}}} \!\!\sum_{(J_j, R_j) \in \mathcal{J}_j} \left(\prod_{j=1}^l \frac{1}{(n \beta)^{P_i/2}} (A)_{J_i}\right) E \left[ \prod_{j=1}^l  ((Y)_{R_i}- E[(Y)_{R_i}] ) \right]~.
\end{eqnarray}

Denoting by $q:= \sum_{i=1}^l P_i/2$, we obtain that 
\begin{eqnarray} \label{trei3}
\!\!\!\!\!\!\!\!E[(\omega_k(n))^l] \!\!&= & \!\!\!\! \sum_{\mbox{{\small $\begin{array}{c} \mathcal{I}_j \in \mathcal{S}_{n,k} \\ 1 \leq j \leq l \end{array}$}}} \!\!\sum_{(J_j, R_j) \in \mathcal{J}_j} \frac{1}{(n \beta)^{q}}\left(\prod_{j=1}^l  (A)_{J_i}\right) E \left[ \prod_{j=1}^l  ((Y)_{R_i}- E[(Y)_{R_i}] ) \right]~.
\end{eqnarray}

%We make two observations: \begin{itemize} \item The non-zero terms in \eqref{trei3} have $q \geq l/2$. Indeed, if that were

\begin{lemma} \label{lemma_q}
The non-zero terms in \eqref{trei3} have $q \geq l/2$. \end{lemma}

\begin{proof} If any of the $l$ terms in the product $ \prod_{j=1}^l  ((Y)_{R_i}- E[(Y)_{R_i}] )$ involves only variables that are independent from all other variables appearing in the remaining $l-1$ terms of the product, the expected value of the product is $0$. This includes the case when at least of the $R_i$s are empty. Hence, in all of the terms that have non-zero contribution to the expectation, each $R_i$ must be nonempty, hence $2q = \sum_{i=1}^l P_i \geq l$. 
\end{proof}

\begin{remark} \label{bound_q} Note that in fact something stronger follows, namely, that for each $1 \leq i \leq l$, there exists an $1 \leq j \leq n$, $j \neq i$, such that the some variable appearing in $(Y)_{R_i}$ also appears in $(Y)_{R_j}$. \end{remark}

Since the entries of $A$ are $O(1)$, and $k$ and $l$ are fixed, the factors $\prod_{j=1}^l  (A)_{J_i}$ are all $O(1)$. By Remark \ref{bound_q} and Lemma \ref{lemma_q}, it follows that every term with non-zero contribution in the double sum \eqref{trei3} corresponds to an $l$-tuple $(R_1, \ldots, R_l)$ with the property described in Remark \ref{bound_q}. Each such $l$-tuple comes with a weight proportional to $1/n^{q}$.

To compute the asymptotics of the sum, we will do the following thought experiment: select a non-zero contribution term and draw the ``correlation'' graph with $R_i$ as vertices, and an edge between $R_i$ and $R_j$ if and only if $(Y)_{R_i}$ and $(Y)_{R_j}$ are correlated. The resulting graph will have $s$ connected components, with $1 \leq s \leq \lfloor \frac{l}{2} \rfloor$. 

Call these connected components $C_1$, \ldots, $C_s$, and consider the set of variables $V_1, \ldots, V_s$, such that $v \in V_i$ if and only if there is an $R_j$ in $C_i$ such that $v$ appears in $(Y)_{R_j}$. Select from each $V_i$ a single variable; these variables will be independent. A variable corresponds to a choice of $1$ index and $3$ possibilities (since it will be of the form $Y_{i, i+1}$, $Y_{i+1, i}$, or $Y_{i,i}$).

If we were to choose a set of $s$ independent variables from $Y$, roughly, to how many such $l$-tuples $(R_1, \ldots, R_{l})$ would this choice correspond, and in turn, to how many sequences $\mathcal{I}$ do these correspond? In other words, to how many non-zero contribution terms in the sum \eqref{trei3} can a choice of $s$ independent variables correspond?

The answer is $O(1)$. 

Indeed, by the way we defined $\mathcal{I}$ and $R$, it follows that, once we have chosen a variable $v \in V_i$, for all other variables in $V_i$ we have a finite number of corresponding indices to choose from. Indeed, this happens because the correlation of $R_i$s induces a ``clustering'' of variables (since all indices must be within $|V_i| \times (k+1) \leq l(k+1)$ of each other). 

Hence, for each of the possible $O(n^s)$ choices of $s$ ``representative'' variables, we have only $O(1)$ possible non-zero contribution terms in the sum \eqref{trei3}.

Going backwards, it follows that for any $s$, there are $O(n^s)$ terms for which the correlation graph has $s$ components. Since each of these terms has weight at most $1/n^{q} \leq 1/n^{l/2}$, we have proved the following lemma.

\begin{lemma} \label{red1} The contribution to the expectation sum \eqref{trei3} from all terms with $s<l/2$ or $q>l/2$ is asymptotically negligible. \end{lemma}

Thus, the only terms of asymptotical significance are those for which $s = q = l/2$. If $l$ is odd, this immediately implies

\begin{lemma} With the notations above, for $k$ and $l$ fixed, $l$ odd,
\[
\lim_{n \rightarrow \infty} E[\omega_k(n)^{l}] = 0~.
\]
\end{lemma}

Let us examine what happens when $l$ is even and $s=q=l/2$. Such terms are easy to understand: they correspond precisely to $l$-tuples $(R_1, \ldots, R_l)$ for which $|R_i| = 1$ for all $i$, and for each $1 \leq i \leq l$ there exists a unique $1 \leq j \leq l$ such that $(Y)_{R_i} = (Y)_{R_j}$.

We make the following simple observation.

\begin{lemma} The number of diagonal terms $Y_{j,j}$ contained in each $(Y)_{R_i}$, counting multiplicities, has to have the same parity as $k$. \end{lemma}
\begin{proof} Indeed, by the definition of any $R$, all the diagonal terms found in $(\tilde{H}_{\beta,n})_{\mathcal{I}}$ must be found in $(Y)_{R}$. The parity of these terms, counting multiplicities, has to be the same as the parity of $k$. This is easy to see; if $\mathcal{I} = (i_1, i_2, \ldots, i_k)$, then by Definition \ref{j_r}\[
i_1-i_2+i_2 - i_3 + \ldots +i_{k-1}-i_{k} +i_{k}- i_1 = 0~,\]and since each difference $i_j - i_{j+1}$ above is either $0$, $1$, or $-1$, it follows that the number of differences equal to $0$ has the same parity as $k$. The number of differences equal to $0$ is the number of diagonal terms. \end{proof}

It then follows that, for all $l$-tuples of $(R_1, \ldots, R_l)$ for which $s = q = l/2$, \begin{itemize}
\item if $k$ is odd, all variables present in the $(Y)_{R_i}$s are diagonal variables, and
\item if $k$ is even, all variables present in the $(Y)_{R_i}$s are off-diagonal variables.
\end{itemize}

We summarize here what we now know about the terms we need to study when $l$ is even.

\begin{lemma} \label{char} The only asymptotically relevant terms have the property that there exist $l/2$ distinct indices $i_1, \ldots, i_{l/2}$ such that for each $i_j$ there exist precisely two values $j_1<j_2$ for which
\begin{enumerate}
\item if $k$ is odd, $R_{j_1} = R_{j_2}= \{(i_j, i_j)\}$.
\item if $k$ is even, either one of these four possibilities:
\begin{itemize}
\item $R_{j_1} = R_{j_2} = \{(i_j, i_j+1)\}$, or 
\item $R_{j_1} = R_{j_2} = \{(i_j+1, i_j)\}$, or 
\item $R_{j_1} = \{(i_j, i_j+1)\}$ and $R_{j_2} = \{(i_j+1, i_j)\}$, or
\item $R_{j_1} = \{(i_j+1, i_j)\}$ and $R_{j_2} = \{(i_j, i_j+1)\}$.
\end{itemize}
Note that in this case, $(Y)_{R_1} = (Y)_{R_2}$, because the matrix is symmetric.
\end{enumerate}
We call all such terms \emph{significant}.
\end{lemma}

We will now need a stronger result than Remark \ref{boundedness}.

\begin{lemma} \label{corners1} For any given $\epsilon>0$ and $k,l \in \mathbb{N}$, with $l$ even, there exists some $i_{\epsilon} \in \mathbb{N}$ such that for any significant term $\left(\prod_{j=1}^l  (A)_{J_i}\right) E \left[ \prod_{j=1}^l  ((Y)_{R_i}- E[(Y)_{R_i}] ) \right]$ \emph{and} the corresponding $l/2$-tuplet $(i_1, \ldots, i_{l/2})$, if $k \leq i_1, \ldots, i_{l/2} \leq n-i_{\epsilon}$, then 
\begin{itemize}
\item if $k$ is odd,
\[
\left | \left(\prod_{j=1}^l  (A)_{J_i}\right) E \left[ \prod_{j=1}^l  ((Y)_{R_i}- E[(Y)_{R_i}] )\right ] - \prod_{m=1}^{l/2} \left( 1- \frac{i_m}{n} \right)^{k-1} ~ \frac{1}{2^{kl}} \right| < \epsilon~;
\]
\item if $k$ is even,
\[
\left | \left(\prod_{j=1}^l  (A)_{J_i}\right) E \left[ \prod_{j=1}^l  ((Y)_{R_i}- E[(Y)_{R_i}] )\right ] - \prod_{m=1}^{l/2} \left(1- \frac{i_m}{n} \right)^{k-1} ~ \frac{1}{2^{kl +l}} \right| < \epsilon~.
\]
\end{itemize}
\end{lemma}

\begin{proof} The lemma follows easily from Proposition \ref{chi_simple} and Remark \ref{convergence}, together with the fact that if $R_{i}$ contains the index $j$, then all indices present in $J_i$ are within $k$ of $j$.
\end{proof}

We prove now that it is enough to look at the significant terms for which
$n-i_{\epsilon} \geq i_1, \ldots, i_{l/2} \geq k$ (i.e., those covered by Lemma \ref{corners1}).

\begin{lemma} \label{corners2}The contribution of significant terms for which some $i_j
>n-i_{\epsilon}$ or $i_j<k$ is asymptotically negligible, i.e. $o(1)$. 
\end{lemma} 
\begin{proof} Each contribution from a significant term \[  
\left ( \prod_{i=1}^l (A)_{J_{i}} \right ) E \left [ \prod_{i=1}^l
((Y_{R_i} - E[(Y)_{R_i}]) \right] \] is bounded by some constant
$\tilde{M}$, by Lemmas \ref{boundedness} and \ref{convergence}. Since
restricting a choice of $i_j$ to be greater than $n-i_{\epsilon}$ or less than
$k$ yields a finite number of choices for that particular $i_j$, and   
since $j<l$ is finite, it follows that there are only $O(n^{l/2-1})$ such
restricted terms. But since the contribution of any such term is weighed
by $1/n^{l/2}$, the statement of the lemma follows. \end{proof}

So we have reduced the computation to examining the contribution from the
terms for which $n-i_{\epsilon} \geq 1_, \ldots, i_{l/2} \geq k$. Assume
w.l.o.g $i_{\epsilon}>k$ (we can always choose a smaller $\epsilon$).

Given an ordered $l/2$-tuplet of distinct indices $n-i_{\epsilon} \geq 1_,
\ldots, i_{l/2} \geq k$, how many terms can correspond to them? First,
there are $(l-1)!!$ ways of pairing these indices to the $R_i$s in this
order. Second, once the pairing is given, \begin{itemize}
 \item for $k$ odd, such a term must have the corresponding
$\mathcal{I}_j$ sequence be a sequence where all but one consecutive    
difference are $\pm 1$ (the one difference that is $0$ corresponds to the
insertion of the diagonal term). There are $\left(k {k-1 \choose
\frac{k-1}{2}}\right)$ such choices for each $\mathcal{I}_j$, for a total of $\left(k {k-1 \choose
\frac{k-1}{2}}\right)^{l}$ choices. 
\item
for $k$ even, such a term have the corresponding $\mathcal{I}_j$ sequence
be a sequence where all consecutive difference are $\pm 1$, and one of
these differences corresponding to the ``marked'' term that belongs to
$R_j$. Taking into account all the 4 possible case, we obtain a total
number of $k^2 {k \choose \frac{k}{2}}^2$ for each pair of matched
$\mathcal{I}_j$s, and total number of $\left (k {k \choose \frac{k}{2}} \right )^{l/2}$ choices. \end{itemize} Note that in either one of the two cases
above, all choices of sequences are valid, because the indices in each
sequence will stay between $1$ and $n$ (this is where we need that all $
k \leq i_j \leq n-i_{\epsilon}$).

Thus, the total number of significant terms which correspond to a given  
ordered $l/2$-tuplet of distinct indices $n-i_{\epsilon} \geq 1_, \ldots,  
i_{l/2} \geq k$ ) is \begin{itemize} \item $(l-1)!! \left(k {k-1
\choose \frac{k-1}{2}}\right)^{l}$ if $k$ is odd, and \item $(l-1)!!
\left(k {k \choose \frac{k}{2}}\right)^{l}$if $k$ is even. \end{itemize}

From Lemmas \ref{red1}, \ref{char}, \ref{corners1} and \ref{corners2}, we obtain 
that for any given $\epsilon$, if $k$ is odd, 
\begin{eqnarray*}
E[(\omega_k(n))^l] & = & \!\!\!\!\!\!\!\!\!\!\!\!\sum_{\begin{array}{c}\mbox{{\small all significant terms}} \\ \mbox{{\small with}}~ k \leq i_j \leq n-i_{\epsilon}, ~\forall ~j  \end{array}} \!\!\!\!\!\!\!\!\!\!\!\!\frac{1}{(n
\beta)^{q}}\left(\prod_{i=1}^l (A)_{J_i}\right) E \left[ \prod_{j=1}^l  
((Y)_{R_i}- E[(Y)_{R_i}] ) \right] +o(1)~, 
\end{eqnarray*} 
and so
\begin{eqnarray*} 
\left | E[(\omega_k(n))^l] -(l-1)!! \left(k {k-1 \choose \frac{k-1}{2}} \right)^{l} \!\!\!\!\!\!\!\!\!\!\!\!\sum_{\begin{array}{c}\mbox{{\small all significant terms}} \\ \mbox{{\small with}}~ k \leq i_j \leq n-i_{\epsilon}, ~\forall ~j \end{array}} \!\!\!\!\!\!\!\!\!\!\!\! \frac{1}{(n \beta)^{l/2}}
\prod_{j=1}^{l/2} \left(1 - \frac{i_j}{4n} \right)^{k-1} \right | && \leq \\ 
 \leq ~(l-1)!! \left(k {k-1 \choose \frac{k-1}{2}}\right)^{l}
\sum_{\begin{array}{c}n-i_{\epsilon} \geq i_1, \ldots, i_{l/2} \geq k \\ \mbox{all} ~i_j~\mbox{distinct} \end{array}} \frac{1}{(n   
\beta)^{l/2}} ~\epsilon ~+o(1)~, & & \\ 
=~ (l-1)!! \left(k
{k-1 \choose \frac{k-1}{2}}\right)^{l} \epsilon (1+o(1)) +o(1)~,~~~~~~~~~~~~~~~~~~~~~~~~~~~~~~~~&& \\
 =~(l-1)!! \left(k {k-1 \choose \frac{k-1}{2}}\right)^{l}
~~~\epsilon ~~(1+o(1))~. ~~~~~~~~~~~~~~~~~~~~~~~~~~~~~~~~~~~~&&
\end{eqnarray*} 
Since $\epsilon$ was arbitrarily
small, it follows that if we can compute 
\[ 
S = (l-1)!! \left(k {k-1 \choose \frac{k-1}{2}}\right)^{l} \!\!\!\!
\sum_{\begin{array}{c} n-i_{\epsilon} \geq i_1, \ldots, i_{l/2} \geq k\\ \mbox{all} ~i_j~\mbox{distinct} \end{array}} \frac{1}{(n
\beta)^{l/2}} \prod_{j=1}^{l/2} \left( 1- \frac{i_j}{4n} \right)^{k-1}~, 
\]  
we are done. But, since $l$ and $k$ are fixed, the sum in $S$ is asymptotically the same as the value of the integral $\left(\frac{1}{\beta}\int_{0}^{1} \left(\frac{1-x}{4}\right)^{k-1}~dx\right)^{l/2}$, hence 
\begin{eqnarray*}
\sum_{\begin{array}{c} n-i_{\epsilon} \geq i_1, \ldots, i_{l/2} \geq k\\ \mbox{all} ~i_j~\mbox{distinct} \end{array}} \frac{1}{(n \beta)^{l/2}}
\prod_{j=1}^{l/2} \left(1-\frac{i_j}{4n} \right)^{k-1} &\sim&
\frac{1}{2^{(k-1)l}} \frac{1}{\beta^{l/2}} \frac{1}{k^{l/2}}~, 
\end{eqnarray*} 
hence 
\[ 
E[(\omega_k(n))^l] - \left(\frac{2}{\beta}\right)^{l/2} \frac{1}{2^{kl}} ~k^{l/2} ~{k-1 \choose \frac{k-1}{2}}^l =O(\epsilon)~,
\]
for arbitrarily small $\epsilon$. 

Similarly, for $k$ even, we obtain through the same sort of calculation that 
\[
E[(\omega_k(n))^l] - \left(\frac{2}{\beta}\right)^{l/2} \frac{1}{2^{(k+1)l}} ~k^{l/2} ~{k \choose \frac{k}{2}}^l =O(\epsilon)~,
\]
for arbitrarily small $\epsilon$. 

Claim \ref{claim1} is thus proved. \qed

\vspace{.25cm}

\noindent \textit{Proof of Claim \ref{claim2}.} The proof is based on the same idea as the proof of Claim \ref{claim1}; the same reasoning applies to yield the asymptotical covariance result. \qed

\begin{remark}
Note that we never actually used the full power of the fact that the entries of the tridiagonal symmetric matrix $Y$ tend to independent centered \textbf{normal} variables. We only used the following three properties: \begin{itemize}
\item $E[Y_{i,i}] = E[Y_{i+1, i}] = 0$;
\item Var$[Y_{i,i}] = \frac{1}{2}$, while $\lim_{n \rightarrow \infty}$ Var$[Y_{i+1, i}] = \frac{1}{8}$;
\item for any $k$, there exists a number $M_k>0$ such that $|E[(Y_{i,j})^k]|<M_k$, for all $1 \leq i,j \leq n$ (boundedness of moments).
\end{itemize}
\end{remark}

\subsection{The $\beta$-Laguerre case} \label{gaus_lag}

Given an integer $k$, consider the random variable
\[
\eta_{k, \gamma}(n) = \mbox{tr}((\tilde{L}_{\beta,n}^{a})^k) - E_{\beta, a}^{L}[\mbox{tr}((\tilde{L}_{\beta,n}^{a})^k)]~.
\]

The main results of this section are given in the Claims below.

\begin{Claim} \label{claim3}
For any fixed integers $k$ and $l$,
\[
\lim_{n \rightarrow \infty} E[(\eta_{k, \gamma}(n))^l] = \left \{ \begin{array}{ll} \left(\frac{2}{\beta}\right)^{l/2} (Sum_1(k, \gamma)+Sum_2(k, \gamma))^{l/2}~(l-1)!!~, & \mbox{if $l$ is even}~,\\
~~0~,& \mbox{if $l$ is odd} \end{array} \right .
\]
where
\begin{eqnarray*}
\!\!\!Sum_1(k, \gamma) &=& \sum_{q=1}^{2k-1} (-1)^{q+1} \gamma^{2k-q} \frac{{2k \choose q}}{2k} \sum_{j=q+1}^{2k} \frac{(-1)^j}{{2k-1 \choose j-1}} \!\!\!\!\sum_{\begin{array}{c} s_1+s_2 = j \\ 1 \leq s_1,s_2 \leq k \end{array}} \!\!\!\! s_1s_2 {k \choose s_1}^2 {k \choose s_2}^2~, \\
\!\!\!Sum_2(k, \gamma) &=& \sum_{q=0}^{2k-2} (-1)^{q} \gamma^{2k-q} \frac{{2k \choose q}}{2k} \sum_{j=q}^{2k-2} \frac{(-1)^j}{{2k-1 \choose j}}  \!\!\!\!\sum_{\begin{array}{c} s_1+s_2 = j \\ 0 \leq s_1,s_2 \leq k\!-\!1 \end{array}}  \!\!\!\!(k-s_1)(k-s_2) {k \choose s_1}^2 {k \choose s_2}^2~.
\end{eqnarray*}
\end{Claim}

\begin{Claim} \label{claim4}
For any fixed integers $k$ and $l$,
\[
\lim_{n \rightarrow \infty} \mbox{Cov}(\eta_{i, \gamma}(n), \eta_{j, \gamma}(n)) =  \frac{2}{\beta} (Sum_1(i, j, \gamma)+Sum_2(i, j, \gamma))~,
\]
where
\begin{eqnarray*}
Sum_1(i, j, \gamma) \!\!\!&=&\!\!\!\! \sum_{q=1}^{i+j-1} (-1)^{q+1} \gamma^{i+j-q} \frac{{i+j \choose q}}{i+j} \sum_{j=q+1}^{i+j} \frac{(-1)^j}{{i+j-1 \choose j-1}} \!\!\!\!\sum_{\begin{array}{c} r+s = j \\ 1 \leq r \leq i \\ 1\leq s \leq j \end{array}} \!\!\!\! rs {i \choose r}^2 {j \choose s}^2~, \\
Sum_2(i, j, \gamma) \!\!\!&=&\!\!\!\!\! \sum_{q=0}^{i+j-2} (-1)^{q} \gamma^{i+j-q} \frac{{i+j \choose q}}{i+j} \sum_{j=q}^{i+j-2} \frac{(-1)^j}{{i+j-1 \choose j}}  \!\!\!\!\!\!\!\!\sum_{\begin{array}{c} r+s = j \\ 0 \leq r \leq i\!-\!1 \\  0 \leq s \leq j\!-\!1 \end{array}}\!\!\!\!\!\!\!\!\!\!(i\!-\!r)(j\!-\!s) {i \choose r}^2 {j \choose s}^2~.
\end{eqnarray*}
\end{Claim}

The method we employ for proving these claims is basically the same as in Section \ref{gaus_herm}; the only things that change are the details of the sequences we will deal with, which in turn effect a change in the calculations, and yield the results of Claims \ref{claim3} and \ref{claim4}. In the following we will point out where definitions and calculations differ from before, but we will not go over the reduction arguments again, for the sake of brevity.

We write the scaled matrix $\tilde{B}_{\beta,n}^{a}$ as
\[
\tilde{B}_{\beta,n}^{a} = D +\frac{1}{\sqrt{\beta n}} Z~,
\]
where $D = E_{\beta, a}^{L}[\tilde{B}_{\beta,n}^{a}]$ is the bidiagonal matrix of mean entries $D(i,i) = \frac{\sqrt{\gamma}}{\sqrt{n \beta}}E[\chi_{2a-i\beta}]$ and $D(i+1, i) =  \frac{\sqrt{\gamma}}{\sqrt{n \beta}}E[\chi_{(n-i)\beta}]$, and $Z$ is the lower bidiagonal matrix of centered variables; we drop the dependence of $D$ and $Z$ on $\beta, a$, and $n$, for simplicity.

\begin{remark} From Proposition \ref{chi_simple}, if $D = (d_{ij})_{1 \leq i,j\leq n}$, given any $\epsilon>0$, there is an $i_{\epsilon} \in \mathbb{N}$ such that
\begin{eqnarray*}
|\sqrt{n \beta/\gamma} ~d_{i,i} - \sqrt{n\beta/\gamma - i \beta}| &\leq& \epsilon~, ~\mbox{and}\\
|\sqrt{n \beta /\gamma} ~d_{i+1,i} - \sqrt{(n-i) \beta}|&\leq& \epsilon~,
\end{eqnarray*}
for any $i \leq n - i_{\epsilon}$.

Here we also used the fact that $2a/(n \beta) \sim 1/\gamma$.

Similarly, again from Proposition \ref{chi_simple}, we know that $Z_{i, i} \rightarrow N(0, \gamma/2)$ and $Z_{i+1,i} \rightarrow N(0,\gamma/2)$ in distribution.
\end{remark}
%\section{Fluctuations: covariance matrix for the monomials} \label{simple_cov 
As in Section \ref{gaus_herm}, we start from the expression for tr$((BB^{T})^k)$, for $B$ a lower bidiagonal matrix:
\[
\mbox{tr}((BB^{T})^k) = \sum_{1 \leq i_1, i_2, \ldots, i_{2k} \leq n} b_{i_1, i_2} b_{i_2, i_3} \ldots b_{i_{2k-1}, i_{2k}} b_{i_{2k}, i_1}~,
\]
where the sum is taken over sequences $(i_1, \ldots, i_{2k})$ with the property that $i_{2j-1}-i_{2j} \in \{0,1\}$, for all $1 \leq j \leq k$, and $i_{2j} - i_{2j+1} \in \{0, -1\}$, for all $1 \leq j \leq k-1$, and also $i_{2k} - i_1 \in \{0, -1\}$.

Just as before, we will introduce a few notations (we ``recycle'' some of the notations we used before; note that the quantities change).

\begin{definition} We denote by $S_{n, k} \in \{1, \ldots, n\}^{2k}$ the set of sequences of integers $i_1, \ldots, i_{2k}$ such that $i_{2j-1}-i_{2j} \in \{0,1\}$, for all $1 \leq j \leq k$, and $i_{2j} - i_{2j+1} \in \{0, -1\}$, for all $1 \leq j \leq k-1$, and also $i_{2k} - i_1 \in \{0, -1\}$. We denote by $\mathcal{I}$ an element in $S_{n, k}$. \end{definition} 

For each such $\mathcal{I}$, we consider all the ways in which we can ``break up'' $\mathcal{I} :=(i_1, \ldots, i_{2k})$ into overlapping ``runs'' $J$ and $R$, i.e.
\[
J = ((i_{p_0}, \ldots, i_{p_1}), (i_{p_2}, \ldots, i_{p_3}), (i_{p_4} \ldots i_{p_5}), \ldots, (i_{p_{2q}}, \ldots, i_{p_{2q+1}}))~,
\] 
and 
\[
R = ((i_{p_1}, \ldots, i_{p_2}), (i_{p_3}, \ldots, i_{p_4}), \ldots, (i_{p_{2q-1}}, \ldots, i_{p_{2q}}))~,
\] 
with 
\[
(i_1, \ldots, i_{2k}) = (i_{p_0}, \ldots, i_{p_1}, i_{p_1+1}, \ldots, i_{p_2}, i_{p_2+1}, \ldots, i_{p_3}, \ldots, i_{p_{2q+1}})~.
\]
Note that this preserves the requirement that $i_{j} - i_{j+1} \in \{0,  (-1)^{1+j \mod 2}\}$ for all $j$.

We allow for the possibility of having empty sequences $i_{p_0}, \ldots, i_{p_1}$ in the beginning and/or $i_{p_{2q}}, \ldots, i_{p_{2q+1}}$ in the end of $J$.

\begin{definition} For any $\mathcal{I}$, we introduce a set $\mathcal{J}$ of pairs $(J,R)$, as described above. For a bidiagonal matrix $B$, we define 
\begin{eqnarray*}
(BB^{T})_{\mathcal{I}} &= &b_{i_1, i_2} b_{i_2, i_3} \ldots b_{i_{2k-1}, i_{2k}} b_{i_{2k}, i_1}~,\\
(BB^{T})_{J} &=&  b_{i_{p_0}, i_{p_0}+1} \ldots b_{i_{p_1}-1, i_{p_1}} b_{i_{p_2} b_{p_2}+1} \ldots b_{i_{p_3}-1, i_{p_3}} \ldots b_{i_{2q}, i_{2q}+1} \ldots b_{i_{2q+1}-1, i_{2q+1}}~; \\
(BB^{T})_{R} &= &b_{i_{p_1}, i_{p_1}+1} \ldots b_{i_{p_2}-1, i_{p_2}} b_{i_{p_3} i_{p_3}+1} \ldots b_{i_{p_4}-1, i_{p_4}} \ldots b_{i_{2q-1}, i_{2q-1}+1} \ldots b_{i_{2q}-1, i_{2q}}~.
\end{eqnarray*}
\end{definition}

\begin{remark} Note that any term in $tr((L_{\beta,n}^{a})^k)$ will consists of terms in $D$ and terms in $Z$, with a sequence of runs $J$ recording the former, and a sequence of runs $R$ recording the latter. \end{remark}

\vspace{.25cm}

\noindent \textit{Proof of Claim \ref{claim4}}. As before, we note that 
\begin{eqnarray} \label{cinci5}
(\tilde{L}_{\beta,n}^a)_{\mathcal{I}} &= & \left(\left(D+\frac{1}{\sqrt{n \beta}} Z \right)\left(D+\frac{1}{\sqrt{n \beta}} Z \right)^{T}\right)_{\mathcal{I}} = \sum_{(J,R) \in \mathcal{J}} \frac{1}{(n \beta)^{P/2}} (D)_{J} (Z)_{R}~,
\end{eqnarray}
with $P = p_2-p_1+1 +\ldots+p_{2q}-p_{2q-1}+1$.

Similarly with \eqref{trei3}, write 
\begin{eqnarray} \label{patru4}
 \!\!\!\!\!\!\!\!E[(\eta_{k, \gamma}(n))^l] \!\!&= & \!\!\!\! \sum_{\mbox{{\small $\begin{array}{c} \mathcal{I}_j \in \mathcal{S}_{n,k} \\ 1 \leq j \leq l \end{array}$}}} \!\!\sum_{(J_j, R_j) \in \mathcal{J}_j} \frac{1}{(n \beta)^{q}}\left(\prod_{j=1}^l  (D)_{J_i}\right) E \left[ \prod_{j=1}^l  ((Z)_{R_i}- E[(Z)_{R_i}] ) \right]~,
\end{eqnarray}
with $q = \sum_i P_i/2$.

The rest of the argument follows in the footsteps of the proof of Claim \ref{claim1}. Just as before, it can be shown that the only terms with significant contribution are those for which, for each $i$, $(Z)_{R_{i}}$ consists of a single term, and, in addition to that, the set of $R_i$s can be split in pairs $(R_{i_1}, R_{i_2})$ such that $(Z)_{R_{i_1}} = (Z)_{R_{i_2}}$. This yields
\[
\lim_{n \rightarrow \infty} E[(\eta_{k, \gamma}(n))^{2p+1}] = 0~.
\]
Also, if $l$ is even, the argument that we can consider only the terms for which there is an $l/2$-tuple $(i_1, \ldots, i_{l/2})$ which ``avoid'' the upper and lower corners of the matrix (like in Lemmas \ref{corners1} and \ref{corners2}) still applies.

The one way in which this computation will differ from the one we made for the proof of Claim \ref{claim1} lies in the fact that approximating $(D)_{J}$, given an index $i_1$ present in $J$, becomes a little trickier, since the diagonal and off-diagonal elements will approximate respectively to $\sqrt{\gamma}\sqrt{1/\gamma - \frac{i}{n}}$ and $\sqrt{\gamma}\sqrt{1 - \frac{i}{n}}$.

Thus, one more parameter will become important, namely, the number of off-diagonal terms in each $(D)_{J_i}$. Note that in each sequence $\mathcal{I}$ we must have an \emph{even} number $2s$ of off-diagonal terms (either from $D$ or from $Z$), since $i_1-i_2+\ldots +i_{2k}-i_{1} = 0$, and this also implies that we have an even number $2(k-s)$ of diagonal terms.

\begin{enumerate} \item Suppose we fix the term in $Z_{R_i}$ to be the diagonal term $z_{i_1, i_1}$; to how many sequences $\mathcal{I}$ with a fixed number $2s$ of off-diagonal terms can this correspond? The answer is $2(k-s) {k \choose s}^2$; we have ${k \choose s}$ ways of picking the off-diagonal terms (because of the alternating property), and once those are picked we have $2(k-s)$ choices for the location of $z_{i_1, i_1}$ among the diagonal terms remaining. 

Each such sequence will have asymptotical weight 
\[
(D)_{J} \sim \gamma^{k - 1/2}\left( \frac{1}{\gamma} - \frac{i}{n} \right)^{k-s -1/2} \left(1- \frac{i}{n} \right)^s~.
\] 

\item Suppose we now fix the term in $Z_{R_i}$ to be the off-diagonal term $z_{i_1+1, i_1}$; to how many sequences $\mathcal{I}$ with a fixed number $2s$ of off-diagonal terms can this correspond? The answer is $2s {k \choose s}^2$; we have ${k \choose s}$ ways of picking the off-diagonal terms (because of the alternating property), and once those are picked we have $2s$ choices for the location of $z_{i_1+1, i_1}$ among them.

Each such sequence will have asymptotical weight 
\[
(D)_{J} \sim \gamma^{k - 1/2}\left( \frac{1}{\gamma} - \frac{i}{n} \right)^{k-s} \left(1- \frac{i}{n} \right)^{s-1/2}~.
\]
\end{enumerate}

Finally, using the binomial formula 
\[
\left(\frac{1}{\gamma} - \frac{t}{n}\right)^{s-1} = \sum_{i=0}^{s-1} (-1)^i {s-1 \choose i} \gamma^{-s+1-i} \left(\frac{t}{n} \right)^i~,
\]
and after some processing and use of the Riemann-sum and Beta-function formula
\begin{eqnarray*}
\lim_{n \rightarrow \infty} \frac{1}{n} \sum_{t=0}^n \left(1-\frac{t}{n}\right)^{2k-r-s} \left (\frac{t}{n} \right)^{r+s-1} & = & \int_0^1 (1-x)^{2k-r-s}~ x^{r+s-1}~dx \\
& = & \frac{(2k-r-s)!~(r+s-1)!}{(2k)!}~,
\end{eqnarray*}
combined with all the possible pairings of the $R_i$s (which yields the necessary $(l-1)!!$), we obtain the result of claim \ref{claim3}. \qed

Claim \ref{claim4} has a similar proof. 

\begin{remark}
As in Section \ref{gaus_herm}, we never actually use the full power of the fact that the entries of the bidiagonal matrix $Z$ tend to independent centered \textbf{normal} variables. We only used the following three properties: \begin{itemize}
\item $E[Z_{i,i}] = E[Z_{i+1, i}] = 0$;
\item  $\lim_{n \rightarrow \infty}$ Var$[Z_{i+1, i}] = \lim_{n \rightarrow \infty} $ Var$[Z_{i,i}] = \frac{\gamma}{2}$;
\item for any $k$, there exists a number $M_k>0$ such that $|E[(Z_{i,j})^k]|<M_k$, for all $1 \leq i,j \leq n$ (boundedness of moments).
\end{itemize}
\end{remark}

\section{A more general setting} \label{gen}

We present here a way to generalize the ``non-random matrix + small random fluctuation'' decompositions we have used in Section \ref{gauss} in order to analyze the deviation and fluctuation from the asymptotical law for the eigenvalue density in the case of the Hermite and Laguerre $\beta$-ensembles. 

\subsection{Tridiagonals}

Let $f$ and $g$ be two integrable functions $f, g : [0,1] \rightarrow \mathbb{R}$ such that the integrals $\int_{0}^{1} f^a(x) g^b(x) dx$ exist for all $a, b \in \mathbb{N}$, and such that the quantities
\begin{eqnarray}
m_k := \sum_{r=0}^{\lfloor \frac{k}{2} \rfloor} {k \choose r, r, k-2r} \int_0^1 f^{k-2r}(x) ~g^{2r}(x) ~dx
\end{eqnarray}
are the moments of a (uniquely determined) distribution $\rho$, defined on a compact $[a,b]$. 

For any $n \in \mathbb{R}$, we then consider the $n \times n$ matrix $F_{T}$:
\[
F_{T} = \left (\begin{array}{cccccc} f\left(\frac{n}{n} \right) & g \left(\frac{n-1}{n} \right) & & & & \\
 g \left(\frac{n-1}{n} \right) & f \left(\frac{n-1}{n} \right) & g \left(\frac{n-2}{n} \right) & & & \\
& g \left( \frac{n-2}{n} \right) & \ddots & \ddots& & \\
& & & \ddots & \ddots & g \left( \frac{1}{n} \right) \\
& & & & g \left( \frac{1}{n} \right) & f \left (\frac{1}{n} \right) \end{array} \right)~,
\]
so the diagonal of $F_T$ is an equidiscretization of $f$ with step $1/n$ (with the exception of the endpoint at $0$, which is missing), and the off-diagonal is an equidiscretization of $g$ with step $1/n$ (missing both the endpoint at $1$ and the one at $0$). 

We consider the tridiagonal symmetric matrix 
\[
R_{T} = \left ( \begin{array}{cccccc} x_n & y_{n-1} & & & & \\
				  y_{n-1} & x_{n-1} & y_{n-2} & & & \\
					  & y_{n-2} & \ddots & \ddots & & \\
					  &     & \ddots & \ddots & y_2 & \\
					  &	& 	 & y_2 & x_2 & y_1 \\
					  &     & 	 &     & y_1 & x_1 \end{array} \right)~,
\]
with the variables $x_i$ and $y_j$ being mutually independent and satisfying the following three properties:
\begin{itemize}
\item $E[x_i] = E[y_j] =0$, for all $1 \leq i \leq n, ~1 \leq j \leq n-1$,
\item Var$[x_i]= \sigma^2$, for all $1 \leq i \leq n$, and Var$[y_j] = \eta^2$, for all $1 \leq j \leq n-1$,
\item for all $k$ there exists a $M_k>0$ such that $|E[(x_i)^k]|<M_k$ and $|E[(y_j)^k]|<M_k$, for all $1 \leq i \leq n, ~1 \leq j \leq n-1$.
\end{itemize}

Note that $F_T$ is a non-random matrix, while $R_T$ is a random one.

Now we consider the matrix model
\begin{eqnarray} \label{model_11}
M_{T} = F_{T} + \frac{1}{\sqrt{n}} R_T~.
\end{eqnarray}

We will compute the asymptotical eigenvalue distribution, the first-order deviation from it, and the first-order fluctuation, for the random matrix $M_{T}$.
 
We need two more definitions.

\begin{definition} Let $P$ be a path on the lattice $\mathbb{Z}^2$, starting at $(0,0)$ and ending at $(k, 0$), with up $((x,y) \rightarrow (x+1, y+1))$, down $((x,y) \rightarrow (x+1, y-1))$, and level $((x,y) \rightarrow (x+1, y))$ steps. For each level $j \in \mathbb{Z}$, we define the quantities $a_j(P)$ and $b_j(p)$, as follows:
\begin{eqnarray*}
a_j(P) & := & \# \mbox{of level steps from $j$ to $j$}; \\
b_j(P) & := & \# \mbox{of down steps from $j$ to $j-1$}~.
\end{eqnarray*}
Note that, since the path $P$ ends at $(k,0)$, the number of up steps it takes must always equal the number of down steps it takes. 

Also let 
\begin{eqnarray*} 
\mathcal{P}_{r, k} & := & \{\mbox{paths from $(0, 0)$ to $(k, 0)$ with exactly $r$ down steps}\}, \\
\mathcal{P}_{r, k, i} & := & \{\mbox{paths in $\mathcal{P}_{r, k}$ which descend to, but not below, $y = -i$}\}, \\
\sl{p}_{r, k, i} & := & |\mathcal{P}_{r,k,i}|, \\
\mathcal{P}_{k} & := & \cup_{r=0}^{\lfloor \frac{k}{2} \rfloor} ~\mathcal{P}_{r, k}~.
\end{eqnarray*}
\end{definition}

\begin{theorem} \label{gen_herm}
Let $M_T$ be a matrix from the ensemble defined by \eqref{model_11}, of size $n$, with eigenvalues $(\lambda_1, \ldots, \lambda_n)$, and let $k \geq 1$ be a positive integer. For all $i =1, \ldots, k$, let
\begin{eqnarray*} 
\mu_i & = & \sum_{r=0}^{\lfloor \frac{i}{2} \rfloor} \left ( f(1)^{i-2r} g(1)^{2r} \left ( \sum_{j=0}^{r-1} (r-j) p_{k,j,r}- \frac{1}{2} {i \choose r, r, i-2r} \right) \right .~~~+\\
& & ~~~~~+~ \left .f(0)^{i-2r} g(0)^{2r} \left(\sum_{j=0}^{r-1} (r-j) p_{i, j, r}- \frac{3}{2} {i \choose r, r, i-2r}  \right ) \right ) ~~ \\
 & & ~~~~~-~ \sum_{P \in \mathcal{P}_{r, i}} \sum_{j \in \mathbb{Z}} j a_j(P) ~\int_0^1 f^{i-2r-1}(t) g^{2r}(t) f'(t) dt ~~~-\\
& & ~~~~~-~ 2\sum_{P \in \mathcal{P}_{r, i}} \sum_{j \in \mathbb{Z}} j b_j(P) ~ \int_0^1 f^{i-2r}(t) g^{2r-1}(t) g'(t) dt ~+~ \\
& & ~~~~~+~\sigma^2 \left(\sum_{P \in \mathcal{P}_{r, i}} \sum_{j \in \mathbb{Z}} {a_j(P) \choose 2} \right) \int_0^1 f^{i-2r-2}(t) g^{2r}(t) dt ~~~~~+~\\
& & ~~~~~+~\eta^2 \left(\sum_{P \in \mathcal{P}_{r, i}} \sum_{j \in \mathbb{Z}} {2b_j(P) \choose 2} \right) \int_0^1 f^{i-2r}(t) g^{2r-2}(t) dt ~.
\end{eqnarray*}
Also, for any $1 \leq i \leq k$, let 
\begin{eqnarray*}
X_i & = & tr((M_T)^i) - n \int_{a}^{b} x^i \rho(x)~dx - \mu_i,\\
%~\frac{1}{4^{i/2} \left( \frac{i}{2}+1 \right) } {i \choose \frac{i}{2}} \delta_{(i \!\!\!\!\mod 2), 0} - \left(\frac{2}{\beta} -1 \right) \int_{-1}^{1} t^i \mu_H(t)~dt,\\
& \equiv & \sum_{j=1}^n \lambda_j^i ~-~ n ~\int_{a}^{b} x^i \rho(x)~dx - \mu_i
%\frac{2}{\pi} \int_{-1}^{1} t^i \sqrt{1-t^2} ~dt ~-~ \left(\frac{2}{\beta} -1 \right) \int_{-1}^{1} t^i \mu_H(t)~dt~~.
\end{eqnarray*}
Let $(Y_1, Y_2, \ldots, Y_k)$ be a centered multivariate Gaussian with covariance matrix 
\begin{eqnarray*} \label{cov_mat_gen_herm}
\mbox{Cov}(Y_i, Y_j) \!\!&=& \!\!\sigma^2 \!\!\!\!\!\!\!\!\!\!\!\!\!\!\sum_{\begin{array}{c} 0 \leq r \leq \lfloor \frac{i}{2} \rfloor-1 \\ 0 \leq s \leq \lfloor \frac{j}{2} \rfloor-1 \end{array}} \!\!\!\!\!\!\!\!\!\!\!\!( i- 2r)(j-2s) {i \choose r, r, i-2r} {j \choose s, s, j-2s} \int_{0}^1 f^{A(r,s)}(x) g^{B(r,s)}(x) ~dx ~+ \\ 
\!\!\!\!\!\!\!\!\!&+&\!\!\eta^2 \!\!\!\!\!\!\!\!\!\!\!\sum_{\begin{array}{c} 1 \leq r \leq \lfloor \frac{i}{2} \rfloor \\ 1 \leq s \leq \lfloor \frac{j}{2} \rfloor \end{array}} \!\!\!\!\!\! 4rs {i \choose r, r, i-2r} {j \choose s, s, j-2s} \int_{0}^1 f^{C(r,s)}(x) g^{D(r,s)}(x) ~dx~,
\end{eqnarray*}
where $A(r, s) = i-2r+j-2s - 2$, $B(r,s) = 2r+2s$, $C(r,s)= 2r + 2s-2$, and $D(r,s) = i-2r + j -2s$.
%\left \{ \begin{array}{ll} \frac{1}{2^{i+j}} \frac{2ij}{i+j} {i-1 \choose \frac{i-1}{2}} {j-1 \choose \frac{j-1}{2}}~,& \mbox{if}~~i=j=1 \mod 2~; \\
% \frac{1}{2^{i+j+2}} \frac{2ij}{i+j} {i \choose \frac{i}{2}} {j \choose \frac{j}{2}}~,&\mbox{if}~~i=j=0 \mod 2~; \\
%0~, & \mbox{otherwise.} \end{array} \right .\end{eqnarray}

Then, as $n \rightarrow \infty$,
\[
(X_1, X_2, \ldots, X_k) ~\Rightarrow ~(Y_1, Y_2, \ldots, Y_k)~.
\]
\end{theorem}

%\begin{theorem} \label{gen_herm}
% Moreover, the fluctuations from $\rho$ are described by a Gaussian process on monomials with covariance matrix given by 
%\begin{eqnarray*}
%Cov(x^k, x^l)\!\! & = & 

%The mean of the process is given by

%for every $k \in \mathbb{N}$.
%\end{theorem}

\begin{proof}
A technical but simple calculation in the spirit of the ones performed in Section \ref{gaus_herm} shows that $m_k$ are the moments of the asymptotic level density, via the expansion of tr$(M_T^k)$ as a sum of products of the entries of $M_T$ and separation of the zero-order term. Furthermore, examining the first-order terms in this expansion yields the covariance result, using the same counting techniques as in Section \ref{gaus_herm}.

Computing the first-order deviation from the mean is slightly more complicated, as first-order terms in the expansion come from four sources; we only enumerate these sources here and indicate how they play a part in the total sum. 

We begin with expressing
\begin{eqnarray*}
E[\mbox{tr}(M_T^k)] & = & E\left [\mbox{tr} \left( \left( F_T+\frac{1}{\sqrt{n}} R_T \right)^k \right) \right ]  \\
%&= & \mbox{tr} (F_T^k) + \frac{1}{\sqrt{n}} \mbox{tr} ( \sum_{a+b = k-1} F_T^{a} ~R_T^{} F_T^{b} ) + \frac{1}{n} \mbox{tr} ( \sum_{a+b+c = k-2} F_T^{a} ~R_T^{} F_T^{b}~R_T )+ o(n^{-1}) ~\\
& = & E \left [ \sum_{\mathcal{I} \in S_{n,k}} (M_T)_{\mathcal{I}} \right ] \\
& = & E \left [\sum_{\mathcal{I} \in S_{n,k}} \sum_{(J,R) \in \mathcal{J}} \frac{1}{n^{P/2}}~(F_T)_J ~(R_T)_R~ \right ]~,
\end{eqnarray*}
where we have used the notation of Section \ref{gaus_herm}.
%
%(F_T)_{i_2, i_3} \ldots (F_T)_{i_{k-1} i_k } (F_T)_{i_k i_1} ~~+ \\
%& & +\sum \sum_{a+b = k-1}  (F_T)_{j_1 j_2} \ldots (F_T)_{j_{a-1} j_{a}} (R_T)_{j_{a} j_{a+1}} (F_T)_{j_{a+1} j_{a+2}} \ldots (F_T)_{j_{k} j_1} ~+~,\\
%& & +\sum \sum_{a+b+c = k-2} (F_T)_{j_1 j_2} \ldots (F_T)_{j_{a-1} j_{a}} (R_T)_{j_{a} j_{a+1}} (F_T)_{j_{a+1} j_{a+2}} \ldots (F_T)_{j_{k} j_1} ~+~,\\
%& & + o(n^{-1/2})~,
%\end{eqnarray*}
%with the first sum ranging over all sequences $i_1, \ldots, i_k, i_{k+1} =: i_1$ such that $|i_j - i_{j+1}| \in \{0,1\}$ and $i_j \in \{1, \ldots, n\}$, while the second and third range over all sequences $j_1, \ldots, j_{k+1} =: j_1$ such that $|j_i - j_{i+1}| \in \{0,1\}$ and $j_i \in \{1, \ldots, n\}$.

It is easy to prove, like we did in Section \ref{gaus_herm}, that the zero-order terms are given by the the pairs $(J,R)$ where $R = \emptyset$, that the terms $(R_T)_R$ which contain a single variable (at first power) are annihilated by the expectation (since all variables in $R_T$ are centered), and that the terms where $(R_T)_R$ contains three or more variables (counting multiplicities) do not contribute to the first-order deviation.

To each sequence $\mathcal{I}$ (recall that $\mathcal{I} = (i_1, \ldots, i_k)$ with $|i_j - i_{j+1}| \in \{0,1\}$ for $1 \leq i \leq k-1$ and $|i_k - i_1| \in \{0,1\}$) we associate in a one-to-one fashion, a path from $(0,0)$ to $(k,0)$ taking steps up, down, or level (depending on the nest term being larger, smaller, or equal to the current one). The zero-order terms sum asymptotically to $m_k$ (with the integral being obtained from the Riemann sum, and ${k \choose r, r, k-2r} = |\mathcal{P}_{k}|$).

%In terms of mean, the second sum never contributes anything, since each term contains exactly one centered Gaussian variable. 

Three of the first-order term sources come from those terms that have $|R| = \emptyset$, while the fourth comes from the terms for which $(R_T)_R$ contains a single variable, at the second power. Note also that the terms for which $(R_T)_R$ contains two different variables will be annihilated by the expectation.
\begin{itemize}
\item[\textit{Source 1.}] In the zero-order count, we ignore the fact that at the ``edges'', i.e. upper left corner, corresponding to $i_1 \in \{1, \ldots, \lfloor \frac{k}{2} \rfloor \}$, and lower right corner, corresponding to $i_1 \in \{n -\lfloor \frac{k}{2} \rfloor, \ldots, n\}$, not all paths in $\mathcal{P}_{k,r}$ can appear in the sum. This approximation yield a first-order term which is asymptotically equal to
\[
S_1 := \sum_{r=0}^{\lfloor \frac{k}{2} \rfloor} \left ( \left( f(1)^{k-2r} g(1)^{2r} + f(0)^{k-2r} g(0)^{2r}\right) \left( \sum_{i=0}^{r-1} (r-i) p_{k,i,r} - {k \choose r, r, k-2r} \right) \right)~. 
\]
\item[\textit{Source 2.}] In the zero-order count, we approximate the value of the integral $\int_0^1 f^{k-2r}(x) ~g^{2r}(x) ~dx$ by the Riemann sum $\frac{1}{n} \sum_{i=0}^n f\left( \frac{i}{n} \right)^{k-2r} g \left( \frac{i}{n} \right)^{2r}$. This yields a first-order term asymptotically equal to\footnote{One can for example use the Euler-Maclauren approximation formulas to obtain this.}
\[
S_2:= \sum_{r=0}^{\lfloor \frac{k}{2} \rfloor}\frac{1}{2}{k \choose r, r, k-2r} \left( f(1)^{k-2r} g(1)^{2r} - f(0)^{k-2r} g(0)^{2r} \right )~.
\] 

\item[\textit{Source 3.}] Finally, in the zero-order approximation, we replace $f\left(\frac{i-j}{n}\right)$ and $g\left(\frac{i-j}{n}\right)$ by $f \left ( \frac{i}{n} \right )$, respectively $g \left ( \frac{i}{n} \right )$ for each $j \in \{ -k, \ldots, k\}$. A Taylor series approximation and the counting of terms shows that this way we need to subtract off have a first-order term which is asymptotically
\[
\!\!\!\!\!\!\!\!\!\!\!\!\!\!\!\!\!\!\!\!\!\!\!\!\!\!\!\!\!\!\!\!S_3 := \sum_{r=0}^{\lfloor \frac{k}{2} \rfloor} \sum_{P \in \mathcal{P}_{r, k}} \sum_{j \in \mathbb{Z}} \left (  j a_j(P) ~\int_0^1 f^{k-2r-1}(t) g^{2r}(t) f'(t) dt - 2 j b_j(P) ~ \int_0^1 f^{k-2r}(t) g^{2r-1}(t) g'(t) dt \right) ~.
\]

\item[\textit{Source 4.}] The last source of first-order terms comes from the terms in which $(R_T)_R$ contains a single variable at the second power,  and its contribution is asymptotically equal to 
\begin{eqnarray*} 
S_4 & :=&  \sigma^2 \left(\sum_{P \in \mathcal{P}_{r, k}} \sum_{j \in \mathbb{Z}} {a_j(P) \choose 2} \right) \int_0^1 f^{k-2r-2}(t) g^{2r}(t) dt ~~+~\\
& & ~~~~+~ \eta^2 \left(\sum_{P \in \mathcal{P}_{r, k}} \sum_{j \in \mathbb{Z}} {2b_j(P) \choose 2} \right) \int_0^1 f^{k-2r}(t) g^{2r-2}(t) dt ~.
\end{eqnarray*}
\end{itemize}

Finally, adding $S_1$, $S_2$, $S_3$, and $S_4$ yields the statement of Theorem \ref{gen_herm}.
\end{proof}

\begin{remark}
Note that when $f(x) = \sqrt{x}/2$, $g(x) = 0$ for all $x \in [0,1]$, $\sigma^2 = \frac{1}{2\beta}$, and $\eta^2 = \frac{1}{8 \beta}$, both the level density asymptotics \emph{\textbf{and}} the covariance matrix for the fluctuations are the same as for the $\beta$-Hermite ensemble of Section \ref{gaus_herm}. 

Crucially, the deviation is \emph{\textbf{different}}. The reason is that in the approximation 
\[
E[\frac{1}{2}\chi_{(n-i)\beta}] \sim \frac{1}{2} \sqrt{(n-i) \beta}~,
\]  the next order term is of order $\frac{1}{\sqrt{(n-i) \beta}}$, which plays a part in computing the deviation. 

We remind the reader that we computed the deviation for the $\beta$-Hermite ensembles by using the palindromic property of expectations of trace, thus reducing the problem to computing the deviation for the ``$\beta = \infty$'' case, for which we used Hermite polynomials properties. This allowed us to find the distribution behind the moments of the deviation.
\end{remark}

\subsection{Bidiagonals}

Let $f$ and $g$ be two integrable functions, $f, g, : [0,1] \rightarrow \mathbb{R}$, such that the integrals $\int_0^1 f^a(x)~g^b(x)~dx$ exist for all $a, b \in \mathbb{N}$, and such that the quantities
\[
\tilde{m}_k := \sum_{r=0}^k {k \choose r}^2 \int_0^1 f^{2r}(x)~g^{2(k-r)}(x)~dx
\]
are the moments of a (uniquely determined) distribution $\nu$ defined on $[a,b]$.

For any $n \in \mathbb{N}$, we consider the $n \times n$ matrix $F_B$, defined below:

%$M_L = M_B M_B^{T}$, where 
%\begin{eqnarray}
%M_B = F_B+\frac{1}{\sqrt{n}} R_B~,
%\end{eqnarray}
%and $F_B$ and $R_B$ are described below.
\[
F_B = \left ( \begin{array}{cccc} f\left( \frac{n}{n} \right) & & &\\
			           g\left( \frac{n-1}{n} \right) & f\left(\frac{n-1}{n} \right) && \\
				   & \ddots & \ddots & \\
				   & & g\left( \frac{1}{n} \right) & f \left( \frac{1}{n} \right) \end{array} \right)~~,
\]
in other words the diagonal of $F_B$ has an equidiscretization of $f$ with step $1/n$ (with the exception  of the endpoint at $0$, which is missing), and the subdiagonal is an equidiscretization of $g$ with step $1/n$ (missing both the endpoint at $0$ and the one at $1$).

We next consider the random bidiagonal matrix
\[
R_B = \left ( \begin{array}{cccc}  x_n & & &\\
			           y_{n-1} & x_{n-1} && \\
				   & \ddots & \ddots & \\
				   & & y_{1} & x_{1} \end{array} \right)~~,
\]
where the variables $x_i,~y_j$ are mutually independent and satisfying the following properties:\begin{itemize}
\item $E[x_i] = E[y_j]=0$, for all $1 \leq i \leq n$, $1 \leq j \leq n-1$,
\item Var$[x_i] = \sigma^2$ for all $1 \leq i \leq n$ and Var$[y_j]=\eta^2$, for all $1 \leq j \leq n-1$, 
\item for all $k$ there is a constant $M_k>0$ such that $|E[(x_i)^k]|<M_k$ and $|E[(y_j)^k]|<M_k$ for all $1 \leq i \leq n$ and $1 \leq j \leq n-1$.
\end{itemize}

Finally, consider the matrix 
\begin{eqnarray} \label{L_mat}
M_L = M_B M_B^T~,
\end{eqnarray}
with 
\[
M_B = F_B + \frac{1}{\sqrt{n}} R_B~.
\]

Note that while $F_B$ is a non-random matrix, $R_B$, $M_B$, and $M_L$ are random.

We will compute the asymptotical level density and the first-order deviation and fluctuation for the random matrix $M_L$.

%A calculation similar to the one we performed in Section \ref{lag_cov}, by expanding the trace of powers of $M_L = M_B M_B^{T}$ as sums of products of entries, and then examining the zero and first order terms and taking the limit as $n \rightarrow \infty$ yields the following Theorem.

We need to define the following quantities.

\begin{definition} Let $Q$ be a path on the lattice $\mathbb{Z}^2$, starting at $(0,0)$ and ending at $(2k, 0$), with up $((x,y) \rightarrow (x+1, y+1))$, down $((x,y) \rightarrow (x+1, y-1))$, and level $((x,y) \rightarrow (x+1, y))$ steps. In addition, we require that the path is alternating, i.e. on each odd-numbered step (first, third, etc.) the path is only allowed to go down or stay at the same level, whereas on each even-numbered step (second, fourth, etc.), the path is allowed only to go up or stay at the same level. 

For each level $j \in \mathbb{Z}$, we define the quantities $c_j(Q)$ and $d_j(Q)$, as follows:
\begin{eqnarray*}
c_j(Q) & := & \# \mbox{of level steps from $j$ to $j$}; \\
d_j(Q) & := & \# \mbox{of down steps from $j$ to $j-1$}~.
\end{eqnarray*}
Note that, since the path $Q$ ends at $(2k,0)$, the number of up steps it takes must always equal the number of down steps it takes. 

Also let 
\begin{eqnarray*} 
\mathcal{Q}_{r, k} & := & \{\mbox{alternating paths from $(0, 0)$ to $(2k, 0)$ with exactly $r$ down steps}\}, \\
\mathcal{Q}_{r, k, i} & := & \{\mbox{alternating paths in $\mathcal{Q}_{r, k}$ which descends to, but not below, $y = -i$}\}, \\
\sl{q}_{r, k, i} & := & |\mathcal{Q}_{r,k,i}|, \\
\mathcal{Q}_{k} & := & \cup_{r=0}^{k} ~\mathcal{Q}_{r, k}~.
\end{eqnarray*}
\end{definition}

\begin{theorem} \label{gen_lag}
Let $M_L$ be the matrix from the ensembles defined by \eqref{L_mat}, of size $n$, with eigenvalues $(\lambda_1, \ldots, \lambda_n)$, and let $k \geq 1$ be a positive integer. For all $1 \leq i \leq k$, let
\begin{eqnarray*} 
\tilde{\mu}_i&= & \sum_{r=0}^{i} \left ( f(1)^{2i-2r} g(1)^{2r} \left ( \sum_{j=0}^{r-1} (r-i) q_{i,j,r}- \frac{1}{2} {i \choose r}^2 \right) \right .~~~+\\
& & ~~~~~+~ \left .f(0)^{2i-2r} g(0)^{2r} \left(\sum_{j=0}^{r-1} (r-j) q_{i, j, r}- \frac{3}{2} {i \choose r, r}^2  \right ) \right ) ~~ \\
 & & ~~~~~-~ \sum_{Q \in \mathcal{Q}_{r, i}} \sum_{j \in \mathbb{Z}} j c_j(Q) ~\int_0^1 f^{2i-2r-1}(t) g^{2r}(t) f'(t) dt ~~~-\\
& & ~~~~~-~ 2\sum_{Q \in \mathcal{Q}_{r, i}} \sum_{j \in \mathbb{Z}} j d_j(Q) ~ \int_0^1 f^{2i-2r}(t) g^{2r-1}(t) g'(t) dt ~+~ \\
& & ~~~~~+~\sigma^2 \left(\sum_{Q \in \mathcal{Q}_{r, i}} \sum_{j \in \mathbb{Z}} {c_j(Q) \choose 2} \right) \int_0^1 f^{2i-2r-2}(t) g^{2r}(t) dt ~~~~~+~\\
& & ~~~~~+~\eta^2 \left(\sum_{Q \in \mathcal{Q}_{r, i}} \sum_{j \in \mathbb{Z}} {2d_j(Q) \choose 2} \right) \int_0^1 f^{2i-2r}(t) g^{2r-2}(t) dt ~.
\end{eqnarray*}
Also, for for any $1 \leq i \leq k$, let 
\begin{eqnarray*}
X_i & = & tr((M_l)^i) - n \int_a^b x^i \nu(x)~dx - \tilde{\mu}_i~, \\
& \equiv& \sum_{j=1}^i \lambda_j^i   - n \int_a^b x^i \nu(x)~dx - \tilde{\mu}_i~
\end{eqnarray*}

Let $(Y_1, Y_2, \ldots, Y_k)$ be a centered multivariate Gaussian with covariance matrix
\begin{eqnarray*}
Cov(x^i, x^j)\!\! & = & \!\!4\sigma^2 \!\!\!\!\!\!\!\!\sum_{\begin{array}{c} 1 \leq r \leq i\\ 1 \leq s \leq j \end{array}} \!\!\!\!\!\!\!\!\! rs {i \choose r}^2 {j \choose s}^2 \int_{0}^1 f^{A(r,s)}(x) g^{B(r,s)}(x) ~dx ~+ \\ 
\!\!\!\!\!\!\!\!\!&+&\!\!4\eta^2 \!\!\!\!\!\!\!\!\!\!\!\sum_{\begin{array}{c} 0 \leq r \leq i-1 \\ 0 \leq s \leq j-1 \end{array}} \!\!\!\!\!\! (i-r)(j-s) {i \choose r}^2 {j \choose s}^2 \int_{0}^1 f^{A(r,s)+2}(x) g^{B(r,s)-2}(x) ~dx~,
\end{eqnarray*}
with $A(r,s) = 2(r+s-1)$ and $B(r,s) = 2(i+j-r-s)$.

Then, as $n \rightarrow n$, 
\[
(X_1, X_2, \ldots, X_k) \Rightarrow (Y_1, Y_2, \ldots, Y_k)~.
\]
\end{theorem}

\begin{proof}
The proof of Theorem \ref{gen_lag} is based on the same calculations as Theorem \ref{gen_herm}. The covariance can be computed using the same general principles as in Section \ref{gaus_lag}, and the examination of the zero-order and first-order terms in the mean can be done as in Section \ref{gen_herm}. Moreover, the sources of the first-order terms are the same as in Section \ref{gen_herm}; it is only the type of path we are counting that changes (from paths of length $k$ to alternating paths of length $2k$).
\end{proof}

\begin{remark} Note that when $f = \sqrt{\frac{1}{\gamma} - 1 + x}, ~g = \sqrt{x}$, $\sigma^2 = \eta^2 = \frac{\gamma}{2 \beta}$,  both the level density asymptotics \emph{\textbf{and}} the covariance matrix for the fluctuations are the same as for the $\beta$-Laguerre ensemble of Section \ref{gaus_lag}. 

Once again, the deviation is \emph{\textbf{different}}, for the same reason as in Section \ref{gen_herm}: in the approximation $E[\frac{1}{2}\chi_{(n-i)\beta}] \sim \frac{1}{2} \sqrt{(n-i) \beta}$  the next order term is of order $\frac{1}{\sqrt{(n-i) \beta}}$, which plays a part in computing the deviation. 

We computed the deviation for the $\beta$-Laguerre ensembles by using the palindromic property of expectations of trace, thus reducing the problem to computing the deviation for the ``$\beta = \infty$'' case, for which we used Laguerre polynomials properties. This allowed us to find the distribution behind the moments of the deviation.
\end{remark}

%\subsection{Generalizing the orthogonal polynomials}

%In this section, we consider a different generalization than in preceding one. In essence, the assumptions will be transferred from the matrix elements to the orthogonal polynomials (with the measure function and inner product) that corresponds to the matrix itself. Given a Jacobi operator with certain properties, we will study the distribution and 

%\subsubsection{Three-term recurrences}

%\subsubsection{Two-term recurrences}

%\section{Extension to a larger (continuous) class of functions} IOANA \& ALAN
\section{Histogramming eigenvalues efficiently} \label{histo}

   We propose a very effective numerical trick for counting
the number of eigenvalues in an interval numerically. 
This method does not require the computation of eigenvalues
and requires a number of operations that is $O(n)$, rather
than $O(n^2)$, which allows for counts for matrices of
a very large size.   
    The method is the standard Sturm sequence method for tridiagonal symmetric 
matrices.  We take as input $D$, a vector of length $n$,
which is the diagonal of the matrix, and $E$, a vector of length $n-1$, the squares of the elements on the super or subdiagonal.  
This avoids unnecessary square roots in the formation of the matrix 
which can slow down computation.

The algorithm is remarkably simple.  For a given value $\sigma$, which is not an eigenvalue of the matrix, compute
$$t_i := D_i - \sigma - E_{i-1}/t_{i-1}$$    
for $t=1$ through $n$ with $t_0=1$ and $E_0=0$.

From Sturm's theorem (see for example \cite[Theorem 13]{gmkrein}) we know that the number of eigenvalues of
the tridiagonal matrix less than or equal to $\sigma$
is the number of $t_i$s that are negative, with $1 \leq i \leq n$.

To histogram eigenvalues of $\beta$-ensembles for $n$'s into the millions or even
the billions, one can simply compute $E$ from a chi-square
random number generator (implemented in MATLAB's Statistics Toolbox as 
$\tt{chi2rnd}$), and $D$ from a normal random number
generator for the $\beta$-Hermite case ($\tt{randn}$) or from another chi-square in the $\beta$-Laguerre case.

An interesting special case is  the $\beta$-Hermite case with $\beta=\infty$, which
computes the roots of the corresponding scaled Hermite polynomial.
This may be performed by taking $D$ to be the zero vector
of length $n$ and $E=(n-1, n-2, \ldots, 2,1)/(4n)$.
 
When $\beta=\infty$, there is no fluctuation, but there are deviations
for large $n$.
Roughly the theorem states that the number of eigenvalues in
an interval $I$ is linear of the form
\begin{eqnarray} \label{form} 
(\# ~\mbox{eigenvalues}) =  n (\mbox{area under the semicircle})  + DEVIATION~,
\end{eqnarray}
where 
\begin{eqnarray} \label{deviation}
DEVIATION = \frac{1}{2\pi}arcsin(x) \Big |_I
\end{eqnarray} 
and we subtract $1/4$ if I contains +1 and $1/4$ if I contains  -1.

\begin{figure}[!ht]
\begin{center}\epsfig{figure = 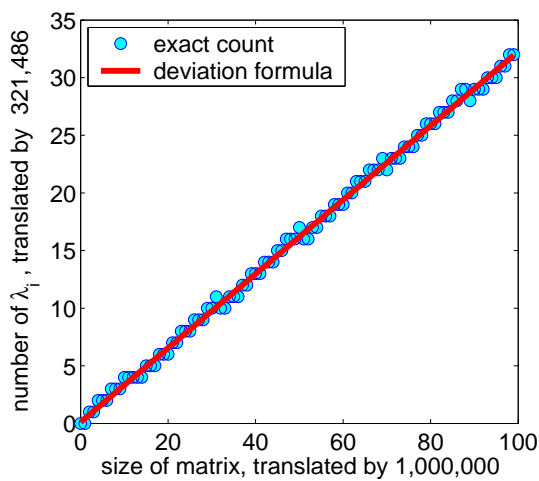, height = 2.5in} \end{center}

\vspace{-.75cm}

\caption{Calculating the deviation in the interval $I = [.2 ~.8]$ for the $\beta = \infty$ Hermite ensemble with sizes $10^6:1:(10^6+99)$; circles represent the eigenvalue count minus the area under the semicircle over $I$; the solid line is the theoretical deviation given in \eqref{deviation}} \label{figura2}
\end{figure}

In one numerical experiment (see Figure \ref{figura2}), we took $n=1,000,000+i$ for $i=0,1,...,99$ and computed the deviation from the mean, i.e.
the number of eigenvalues in the interval $I$ minus the area $ A= n\frac{1}{2 \pi} \int_{I} \sqrt{1-x^2}~dx$;  we did this arbitrarily for the interval $ I = [.2 ~.8]$. Since this experiment is non-random it is repeatable without
any reference to a random number generator.  We found the
experimental deviation of 0.1167 which is close to the theoretical
deviation of .1155, given by \eqref{deviation}.

Figure \ref{figura2} plots the results of the experiment.  The red line
which contains the theoretical value of the intercept represents
the best fit line to the data in the sense that the average
vertical deviation is minimized.

\section{Remarks and open problems} 

There are many object-counting (combinatorial) approaches to the study of traces of powers of random matrices; they depend on the matrix model, and on the polynomial whose trace is being computed. For example, the counting approach of \cite{sosh_sinai2} uses full matrix models (all entries are non-zero variables), and traces of powers (thus using the monomial basis, like we have done here), and counts paths in the complete graph of size $n$. By contrast, in \cite{speicher_new}, the polynomials used are the shifted Chebyshev polynomials, and the objects counted are non-crossing annular partitions; the matrix models are still full. Here, we use tri/bidiagonal matrix models, consider the monomials, and count essentially paths with three types of steps (up, down, level) in the plane.

Though the objects we count here are simpler than in \cite{speicher_new}, our counting technique expresses the results in a less compact form than in \cite{johansson_clt_herm} and \cite{speicher_new}. In the latter two papers, the covariance matrix is diagonalized by the choice of polynomial basis, whereas in our paper it is obtained as full because we work with the monomials. There seems to be a trade-off between the simplicity of the object to be counted and the simplicity of the form in which the covariance matrix is expressed.

We would like to conjecture that by using a hybrid way of counting, for example, using the tridiagonal matrices and some of the techniques of \cite{speicher_new}, both the counting process and the resulting format of the answer could be simplified. The development of such a technique would be of great interest.

\section{Acknowledgments} The authors would like to thank Alexei Borodin, Alexander Soshnikov, and Roland Speicher for many useful conversations on the subjects of global fluctuations, traces of powers of random matrices, and free probability. We also thank Inderjit Dhillon for providing Sturm sequence code, which we used for the histogramming of eigenvalues. 

This research study has been conducted while Ioana Dumitriu was a Miller Research Fellow, sponsored by the Miller Institute for Basic Research in Sciences at U.C. Berkeley. 

Finally, the authors would like to thank the National Science Foundation (DMS-0411962).

\bibliography{bib_10_01}
\bibliographystyle{plain}

\end{document}